\documentclass[a4paper]{article}

\usepackage{amsmath,amstext,amssymb,amsfonts}
\usepackage{color,graphicx}
\usepackage{tabularx}
\usepackage{verbatim}

\usepackage{amsthm}
\usepackage{ifdraft}

\usepackage{nicefrac}

\usepackage{microtype}

\newtheorem{theorem}{Theorem}[section]
\newtheorem*{theorem*}{Theorem}

\newtheorem{Claim}[theorem]{Claim}
\newtheorem*{claim*}{Claim}

\newtheorem{proposition}[theorem]{Proposition}
\newtheorem*{proposition*}{Proposition}
\newtheorem{lemma}[theorem]{Lemma}
\newtheorem*{lemma*}{Lemma}

\newtheorem*{conjecture*}{Conjecture}
\newtheorem{observation}[theorem]{Observation}
\newtheorem{fact}[theorem]{Fact}
\newtheorem*{fact*}{Fact}

\newtheorem*{hypothesis*}{Hypothesis}

\theoremstyle{definition}
\newtheorem{definition}[theorem]{Definition}

\newtheorem{algorithm}[theorem]{Algorithm}
\newtheorem{SDP}[theorem]{SDP}

\newtheorem{remark}[theorem]{Remark}

\usepackage{prettyref}
\newcommand{\savehyperref}[2]{\texorpdfstring{\hyperref[#1]{#2}}{#2}}

\newrefformat{eq}{\savehyperref{#1}{\textup{(\ref*{#1})}}}
\newrefformat{lem}{\savehyperref{#1}{Lemma~\ref*{#1}}}
\newrefformat{def}{\savehyperref{#1}{Definition~\ref*{#1}}}
\newrefformat{thm}{\savehyperref{#1}{Theorem~\ref*{#1}}}
\newrefformat{cor}{\savehyperref{#1}{Corollary~\ref*{#1}}}
\newrefformat{cha}{\savehyperref{#1}{Chapter~\ref*{#1}}}
\newrefformat{sec}{\savehyperref{#1}{Section~\ref*{#1}}}
\newrefformat{app}{\savehyperref{#1}{Appendix~\ref*{#1}}}
\newrefformat{tab}{\savehyperref{#1}{Table~\ref*{#1}}}
\newrefformat{fig}{\savehyperref{#1}{Figure~\ref*{#1}}}
\newrefformat{hyp}{\savehyperref{#1}{Hypothesis~\ref*{#1}}}
\newrefformat{alg}{\savehyperref{#1}{Algorithm~\ref*{#1}}}
\newrefformat{sdp}{\savehyperref{#1}{SDP~\ref*{#1}}}
\newrefformat{rem}{\savehyperref{#1}{Remark~\ref*{#1}}}
\newrefformat{item}{\savehyperref{#1}{Item~\ref*{#1}}}
\newrefformat{step}{\savehyperref{#1}{step~\ref*{#1}}}
\newrefformat{conj}{\savehyperref{#1}{Conjecture~\ref*{#1}}}
\newrefformat{fact}{\savehyperref{#1}{Fact~\ref*{#1}}}
\newrefformat{prop}{\savehyperref{#1}{Proposition~\ref*{#1}}}
\newrefformat{claim}{\savehyperref{#1}{Claim~\ref*{#1}}}
\newrefformat{relax}{\savehyperref{#1}{Relaxation~\ref*{#1}}}
\newrefformat{red}{\savehyperref{#1}{Reduction~\ref*{#1}}}
\newrefformat{part}{\savehyperref{#1}{Part~\ref*{#1}}}
\newrefformat{prob}{\savehyperref{#1}{Problem~\ref*{#1}}}
\newrefformat{ass}{\savehyperref{#1}{Assumption~\ref*{#1}}}

\newcommand{\Sref}[1]{\hyperref[#1]{\S\ref*{#1}}}

\renewcommand{\leq}{\leqslant}

\renewcommand{\geq}{\geqslant}

\usepackage{bm}

\usepackage{xspace}

\usepackage[pdftex,pagebackref,colorlinks,linkcolor=blue,filecolor = blue, citecolor = blue, urlcolor  = blue]{hyperref}

\usepackage[top=2cm, bottom=2cm, left=1.6cm, right=1.6cm]{geometry}

\usepackage{boxedminipage}

\newcommand{\mper}{\,.}
\newcommand{\mcom}{\,,}

\newcommand{\paren}[1]{\left(#1 \right )}

\newcommand{\Brac}[1]{\left[#1\right]}

\newcommand{\set}[1]{\left\{#1\right\}}

\newcommand{\abs}[1]{\left\lvert#1\right\rvert}
\newcommand{\Abs}[1]{\left\lvert#1\right\rvert}

\newcommand{\ceil}[1]{\lceil #1 \rceil}

\newcommand{\norm}[1]{\left\lVert#1\right\rVert}

\newcommand{\defeq}{\stackrel{\textup{def}}{=}}

\newcommand{\Z}{{\mathbb Z}}
\newcommand{\N}{{\mathbb Z}_{\geq 0}}
\newcommand{\R}{\mathbb R}
\newcommand{\Q}{\mathbb Q}

\newcommand{\sdp}{{\sf SDP }}

\newcommand{\OPT}{{\sf OPT}}

\newcommand{\seteq}{\mathrel{\mathop:}=}
\newcommand{\subjectto}{\text{subject to}}

\newcommand{\Esymb}{\mathbb{E}}
\newcommand{\Psymb}{\mathbb{P}}

\DeclareMathOperator*{\E}{\Esymb}

\DeclareMathOperator*{\ProbOp}{\Psymb}

\newcommand{\Ex}[1]{\E\Brac{#1}}

\renewcommand{\Pr}[1]{\ProbOp\Brac{#1}}

\newcommand{\e}{\epsilon}

\definecolor{DSgray}{cmyk}{0,0,0,0.7}

\let\e\varepsilon

\newcommand{\cF}{\mathcal F}

\newcommand{\etal}{et. al.}
\newcommand{\bigO}{\mathcal{O}}
\newcommand{\bigo}[1]{\bigO\left(#1\right)}

\newcommand{\poly}{{\sf poly}}

\newcommand{\U}{\bar{u}}

\usepackage{fullpage}
\usepackage{bbm} 
\usepackage{todonotes}

\usepackage{algpseudocode}
\usepackage{algorithm}
\algnewcommand\algorithmicinput{\textbf{Input:}}
\algnewcommand\INPUT{\item[\algorithmicinput]}
\algnewcommand\algorithmicoutput{\textbf{Output:}}
\algnewcommand\OUTPUT{\item[\algorithmicoutput]}
\algnewcommand\algorithmicparameters{\textbf{Parameters:}}
\algnewcommand\PARAMETERS{\item[\algorithmicparameters]}

\newcommand{\vbmfull}{vertex expansion block model}
\newcommand{\vbm}{{\sf VBM}}

\newcommand{\trace}{{\sf trace}}
\newcommand{\one}{\mathbbm{1}}
\newcommand{\inds}{\one_S}
\newcommand{\phiv}{\phi^{\sf V}}
\newcommand{\eps}{\varepsilon}

\newcommand{\phiopt}{\phiv} 
\newcommand{\phialg}{\phi^{\sf V}_{\mathrm{ALG}}}
\newcommand{\phisdp}{\phi^{\sf V}_{\mathrm{SDP}}}
\newcommand{\phidual}{\Gamma^{\sf V}}

\newcommand{\insquare}[1]{\left[#1\right]}
\newcommand{\inbrace}[1]{\left\{ #1 \right\}}

\newcommand{\diag}{\mathrm{diag}}

\newcommand{\eqdef}{\seteq}
\newcommand{\g}{\one_S} 
\renewcommand{\epsilon}{\e} 

\newcommand{\mP}{\mathcal{P}}

\newcommand{\mD}{\mathcal{D}}
\newcommand{\mF}{\mathcal{F}}

\begin{document}

\title{
Semi-random Graphs with Planted Sparse Vertex Cuts: Algorithms for Exact and Approximate Recovery}

\author{Anand Louis \footnote{\texttt{E-mail: anandl@iisc.ac.in}}\\
  Indian Institute of Science \\ Bangalore, India.
  \and
  Rakesh Venkat\footnote{Supported by an I-Core Algorithms Fellowship. \texttt{E-mail:rakesh@cs.huji.ac.il}}\\
  Hebrew University of Jerusalem \\ Israel}

\date{}

\begin{titlepage}
\maketitle

\begin{abstract}
The problem of computing the vertex expansion of a graph is an NP-hard problem.
The current best worst-case approximation guarantees for computing the vertex expansion of a graph 
are a $\bigo{\sqrt{\log n}}$-approximation algorithm due to Feige \etal~\cite{fhl08}, and 
$\bigo{\sqrt{\OPT \log d}}$ bound in graphs having vertex degrees at most $d$ due to
Louis \etal~\cite{lrv13}.

We study a natural semi-random model of graphs with sparse vertex cuts.
For certain ranges of parameters, we give an algorithm to recover the planted 
sparse vertex cut exactly. For a larger range of parameters, we give a constant factor bi-criteria
approximation algorithm to compute the graph's balanced vertex expansion.
Our algorithms are based on studying a semidefinite programming relaxation for the 
balanced vertex expansion of the graph.

In addition to being a family of instances that will help us to better understand the complexity of
the computation of vertex expansion, our model can also be used in the study of
community detection where only a few nodes from each community interact with nodes from other 
communities. 
There has been a lot of work on studying random and semi-random graphs with
planted sparse edge cuts. 
To the best of our knowledge, our model of semi-random graphs with planted 
sparse vertex cuts has not been studied before.

\end{abstract}
\thispagestyle{empty}

\end{titlepage}

\newcommand{\phivb}{\phi^{\sf V-bal}}
\newcommand{\phiva}{\phi^{\sf V,a}}
\newcommand{\vbmparams}{\vbm$(n,\e_1,\e_2,p_1,p_2,c,r,\lambda_1, \lambda_2)$}
\newcommand{\vbmparamslambda}{\vbm$(n,\e_1,\e_2,p_1,0,0,0,0,0)$}
\newcommand{\vbmparamsp}{\vbm$(n,\e_1,\e_2,0,0,0,r,\lambda_1,0)$}
\newcommand{\randomgraph}{pseudo-random graph}

\section{Introduction}

Given a graph $G = (V,E)$, the vertex expansion of a non-empty subset $S \subset V$, denoted 
by $\phiv(S)$, is defined as\footnote{
Other definitions of vertex expansion have been studied in the literature, see \prettyref{sec:relatedwork}.}
\[  \phiv(S) \defeq \Abs{V} \frac{\Abs{N(S)} + \Abs{N(V \setminus S)} }{ \Abs{S} \Abs{V \setminus S}} \mcom  \]
where $N(S)$, the neighborhood of $S$, is defined as 
$ N(S) \defeq \set{j \in V \setminus S : \exists i \in S \textrm{ such that } \set{i,j} \in E }$.
The vertex expansion of the graph $G$, denoted by $\phiv_G$, is defined as
$\phiv_G \defeq \min_{S \subset V, \ S \neq \emptyset} \phiv(S)$.
Computing the vertex expansion of a graph is NP-hard. 
The complexity of computing various graph expansion parameters are central open problems 
in theoretical computer science, and inspite of many decades of intensive research,
they are yet to be fully understood \cite{a85,am85,lr99,arv09,fhl08,rs10}.

Feige \etal~\cite{fhl08} gave a $\bigo{\sqrt{\log n}}$-approximation algorithm for computing the vertex 
expansion of a graph. Louis \etal~\cite{lrv13} gave an algorithm that computes a set having vertex expansion at most
$\bigo{\sqrt{\phiv \log d}}$ in graphs having vertex degrees at most $d$. 
We give a brief description of other related works in \prettyref{sec:relatedwork}.
In this work, we study a natural semi-random family of graphs,
and give polynomial
time exact and approximation algorithms for computing the
{\em balanced vertex expansion} (a notion that is closely related
to the vertex expansion of a graph, we define it formally in \prettyref{sec:model}) w.h.p.

In many problems, there is a huge gap betwen theory and practice;
the best known algorithms provide a somewhat underwhelming performance guarantee, however simple heuristics
perform remarkably well in practice. 
Examples of this include the simplex algorithm for linear programming \cite{km72},
SAT \cite{bm99}, sparsest cut \cite{kk95,kk98}, among others. 
In many cases, the underwhelming provable approximation guarantee of an algorithm is a property 
(hardness of approximation) of the problem itself; even in
many such cases, simple heuristics work remarkably well in practice.
A possible explanation for this phenomenon could be that for many problems, the instances arising in practice tend
to have some inherent structure that makes them ``easier'' than the worst case instances. Many attempts have been made
to understand the structural properties of these instances, and to use them in designing algorithms specifically for such
instances, which could perform much better than algorithms for general instances.
A fruitful direction of study has been that of 
modelling real world instances as a family of random and semi-random instances satisfying certain properties.  
Our work can be viewed as the study of the computation of vertex expansion along this direction.

Often graphs with sparse cuts are used to model communities. For example, the vertices of a graph can 
be used to represent the members of the communities, and two vertices would have an edge between them if the
members corresponding to them are {\em related} in some way. In such a graph, the sparse cuts
indicate the presence of a small number of relations across the members corresponding to the cut, 
which are likely to be some form of communities within the members.
The {\em stochastic block models} have been used to model such communities. 
Our model can also be viewed as model for communities where only a few members from each community have a relationship 
with members from another community.

\paragraph{Organization.}
We define our model in \prettyref{sec:model}, and state our results in \prettyref{sec:results}.
We give our \sdp~relaxation in \prettyref{sec:sdp}.
We give an overview of our proofs in \prettyref{sec:proofoverview}, and present the proofs 
of our theorems in \prettyref{sec:proof} and \prettyref{sec:const-factor-approx}.

\subsection{Vertex Expansion Block Models.}
\label{sec:model}
For a graph $G = (V,E)$, its {\em balanced vertex expansion} $\phivb$
is defined as
\[ \phivb_G \defeq \min_{ \substack{S \subset V \\ \Abs{S} = \Abs{V}/2 }} \phiv(S) \mper  \]

Another common notion of vertex expansion that has been studied in the literature is  
$\phiva(S) \defeq  \paren{ \Abs{V} \Abs{N(S)}/\paren{ \Abs{S} \Abs{V \setminus S}}}$, 
and as before, $\phiva_G \defeq \min_{S \subset V} \phiva(S)$.
\cite{lrv13} showed that the computation $\phiv_G$ and $\phiva_G$ is equivalent upto constant factors. In this work, we develop a semi-random model for investigating the balanced vertex expansion of graphs.
%
%

We study instances that are constructed as follows. We start with a set of $n$ vertices, and
we arbitrarily partition them into two sets $S$, $S'$ of $n/2$ vertices each. Next, we choose
a small subset $T \subset S$ of size $\e n$ (resp. $T' \subset S'$) to form the vertex boundary of these sets.
On $S \setminus T$ (resp. $S' \setminus T'$), we add an arbitrary graph whose spectral gap\footnote{\label{sg}
The spectral gap of a graph is defined as the second smallest eigenvalue of its normalized Laplacian matrix,
see \prettyref{sec:notation} for definition} 
is at least $\lambda$ (a parameter in this model), and whose vertices have roughly the same degree.
We add an arbitrary low degree bipartite graph between $T$ and $T'$. 
Between each pair of vertices in $(S \setminus T) \times T$, we add edges 
independently at random with probability $p$; this is the only part of the construction that is random. 
Next, we allow a {\em monotone adversary} to alter the graph : the monotone adversary can  
arbitrarily add edges that do not change the sparsity
of the vertex cut $(S,S')$, i.e., add edges between any pair of vertices in $S$
(resp. $S'$), and between any pair in $T \times T'$.

\begin{figure}
\centering
\label{fig:model}
\includegraphics[scale=0.5]{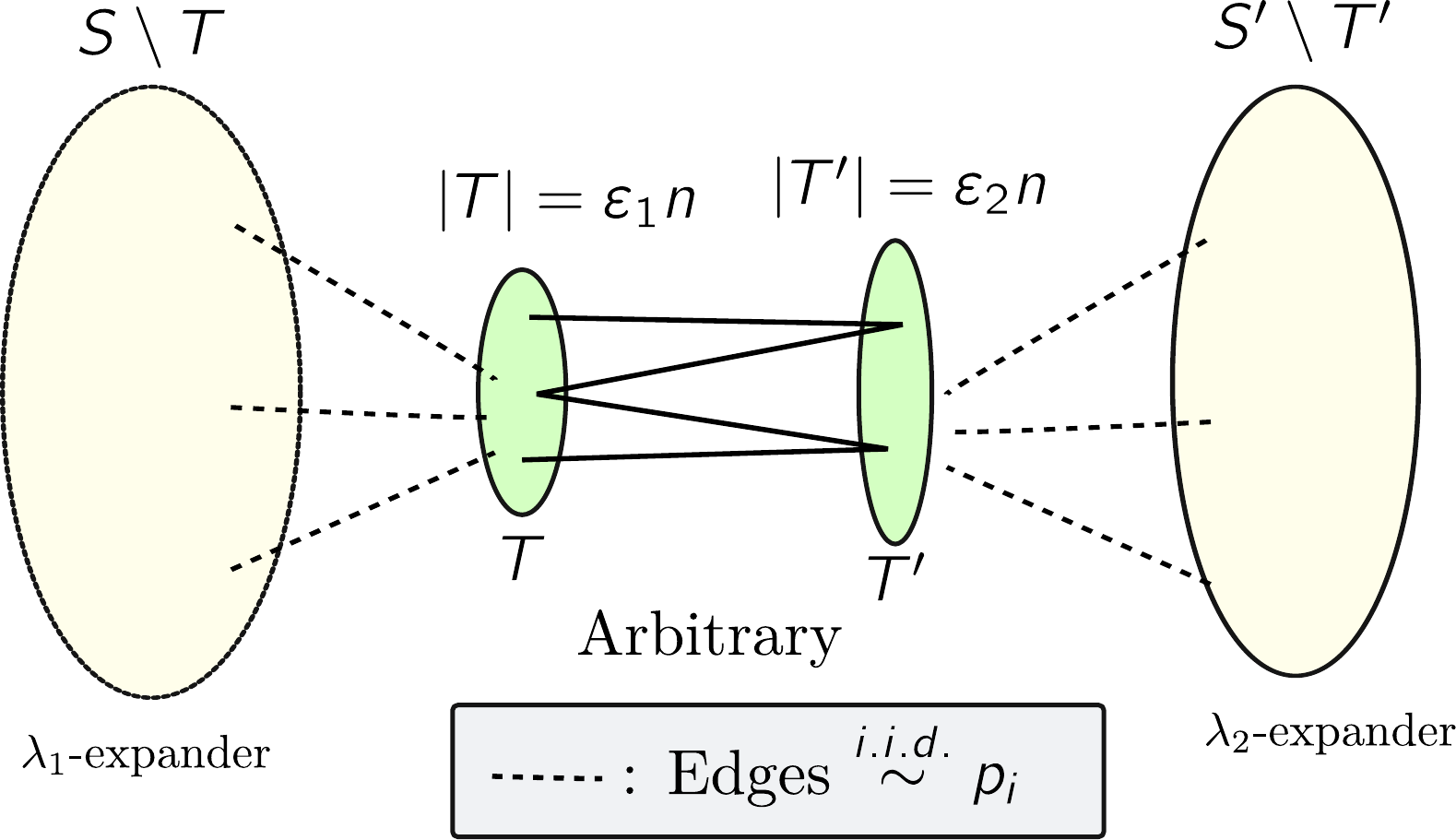}
	\caption{\vbmparams. A monotone adversary may further add arbitrary edges within $S$, $S'$ and between $T, T'$.}
\end{figure}

In our model, we allow the sets $S$ and $S'$ to be generated using different sets of 
parameters, i.e., we use $\e_1, \lambda_1, p_1$ for $S$ and $\e_2, \lambda_2, p_2$ for $S'$. 
We formally define the \vbmfull~below (see also \prettyref{fig:model}); we refer to it as the \vbm~model.

\begin{definition}[The \vbm~model] 
\label{def:vbm} 
An instance of \vbmparams~is generated as follows.
\begin{enumerate}
\item \label{step:part} 
	Let $V$ be a set of $n$ vertices. Partition $V$ into two sets $S$ and $S'$ of $n/2$
	vertices each. Partition $S$ into two sets $T$ and $S \setminus T$ of sizes
	$\e_1 n$ and $(1/2 - \e_1) n$ respectively.
	Similarly, partition $S'$ into two sets $T'$ and $S' \setminus T'$ of sizes
	$\e_2 n$ and $(1/2 - \e_2) n$ respectively.
\item \label{step:rand} 
	Between each pair in $(S \setminus T) \times T$ (resp. $(S' \setminus T') \times T'$), add 
	an edge independently with probability $p_1$ (resp. $p_2$).
\item \label{step:arbs}
	Between pairs of vertices in $S \setminus T$ (resp. $S' \setminus T'$), add edges 
	to form an arbitrary roughly regular (formally, ratio of the maximum vertex degree and the 
		minimum vertex degree is at most $r$) of spectral gap\textsuperscript{\ref{sg}}
		at least $\lambda$.
\item \label{step:tt}
	Between pairs in $T \times T'$, add edges to form an arbitrary bipartite 
	graph of vertex degrees in the range $[1,c]$ (this bipartite graph need not be connected);
	if $c<1$, then add no edges in this step. We will use $\cF$ to denote this bipartite graph.
\item \label{step:madv} ({\em Monotone Adversary}) Arbitrarily add edges between any pair of vertices in $S$ (resp. $S'$).
		Arbitrarily add edges between any pair in $T \times T'$.
\end{enumerate}
Output the resulting graph $G$. 
\end{definition}

We note that the direct analogue for vertex expansion of \emph{Stochastic Block Models} (see related work in Section~\ref{sec:relatedwork}) in the regimes allowing for exact recovery is included in this setting: there, the graphs within $S$ and $S'$ are completely random, and so are the connections between $T$ and $T'$ (before the monotone adversary acts). Our model allows for a lot more adversarial action, while restricting the randomness to only a small portion of the graph.

In addition to being a family of instances that will help us to better understand 
the complexity of the computation of vertex expansion, the \vbmfull~can also be used
in the study of community detection. In the case of two communities, the vertices in 
$S$ and $S'$ can model the members of the communities. Each community can have a few 
representatives who interact with the representatives from other communities;
these representatives can be modelled using $T$ and $T'$, and their interactions
can be modelled by the arbitrary graphs within $T$ and $T'$, and the
low degree bipartite graph and the action of the monotone adversary between $T$ and $T'$.
Even though the connections within a community may be arbitrary,
usually the members within the community are well connected with each other; 
this can be modelled by the choosing an appropriate values of $\lambda_1, \lambda_2$
plus the action of the monotone adversary.
We can model the connections between community members and their representatives 
by a sparse random bipartite graph; our model allows the flexibility of choosing
$p_1$ and $p_2$, and also the action of the monotone adversary.

\subsection{Our Results}
\label{sec:results}
Our main result is a polynomial time algorithm for exactly recovering $S$ and $S'$
from a graph sampled from \vbmparams~for certain ranges of parameters. 
\begin{theorem}
\label{thm:main}
There exist universal constants 
$c_1 \in \R^+, c_2 \in (0, 1/2), c_3 \in \Z^+, c_4 \in \R^+, c_5 \in (0,1)$ satisfying the following: 	
there exists a polynomial time algorithm which takes a graph generated from \vbmparams,
where $p_1, p_2 \in (c_1 (\log n)/n, 1]$, $\e_1,\e_2 \in [1/n, c_2]$,  $c \leq c_3$, $r \leq c_4$
and $\lambda_1, \lambda_2 \geq c_5$,
and outputs the sets $S$ and $S'$ with probabilty at least $1 - 1/{\sf poly}(n)$.
\end{theorem}

We prove \prettyref{thm:main} in \prettyref{sec:proof}; in fact, we prove a slightly more general 
result (\prettyref{thm:sdpintegral-general}).

We also show that if the instances satisfy a few weaker requirements, then we can
obtain a constant factor bi-criteria approximation algorithm for 
computing the balanced vertex expansion. 

We study the case when $S \setminus T$ is an arbitrary graph, i.e., 
it does not have constant spectral gap.
Note that this case is captured by setting $\lambda_1 = 0$ in our model, since the 
monotone adversary can create any arbitrary graph on $S \setminus T$. 
Our proof also allows us to let the graph induced on $S'$ be an arbitrary graph. Again, this is captured by 
setting $p_2 = \lambda_2 = 0$ in our model, since the monotone adversary can create
any arbitrary graph on $S'$. 
We show that we can use the underlying random bipartite graph between $S \setminus T$ 
and $T$ to obtain a constant factor bi-criteria approximation algorithm for computing the 
balanced vertex expansion in this case.

\begin{theorem}
\label{thm:vbmlambda}
There exist universal constants 
$c_1 \in \R^+$, $c_2 \in (0, 1/2), c_6 \in (0,1/2)$ satisfying the following:	
there exists a polynomial time algorithm which takes a graph generated from \vbmparamslambda,
where $\e_1,\e_2 \in [1/n, c_2]$ and $p_1 (\e_1 + \e_2) n \geq c_1 \log n$,
and outputs  with probability at least $1 - 1/{\sf poly}(n)$, 	
a set $A \subset V$ satisfying $\Abs{A} \in [ c_6 n, (1 - c_6) n]$
and $\phiv(A) \leq \bigo{\e_1 + \e_2}$.
\end{theorem}

Next, we study the case where the edges between $S \setminus T$ and $T$ are arbitrary, but
$\lambda_1$ is large.
As in the previous case, our proof allows the graph induced on $S'$ to be an 
arbitrary graph.
Again, as before, this case is captured by setting $p_1 = p_2 = \lambda_2 = 0$.
In this case, we show that for certain ranges of $\lambda_1$, we can 
obtain a constant factor bi-criteria approximation algorithm for 
computing the balanced vertex expansion.

\begin{theorem}
\label{thm:vbmp}
There exist universal constants
$c_2 \in (0, 1/2), c_5 \in (0,1), c_6 \in (0,1/2)$ satisfying the following: 	
there exists a polynomial time algorithm which takes a graph generated from \vbmparamsp,
where $\e_1,\e_2 \in [1/n, c_2]$, and $\lambda_1 \geq c_5 r^3 (\e_1 + \e_2)$,
and outputs 	
a set $A \subset V$ satisfying $\Abs{A} \in [ c_6 n, (1 - c_6) n]$
and $\phiv(A) \leq \bigo{\e_1 + \e_2}$.
\end{theorem}

In \prettyref{sec:vbmp}, we prove a stronger result: it suffices for $S \setminus T$ to contain
a subgraph on $\Omega(n)$ vertices having spectral gap at least $c_5 r^3 (\e_1 + \e_2)$,
to obtain a constant factor bi-criteria approximation algorithm for computing the 
balanced vertex expansion (\prettyref{thm:vbmp-main}).

\subsection{Related Work}
\label{sec:relatedwork}

\paragraph{Stochastic Block Models.}
Closely related to the vertex expansion of a graph is the notion of edge expansion which is defined as follows.
\begin{definition}
\label{def:edgeexpansion}
For a weighted graph $G = (V,E,w)$, with non-negative edge weights $w : E \to \Q^+$, 
the edge expansion of a non-empty set $S \subset V$ is defined as 
\[ \phi_G(S) \defeq \frac{ \sum_{ e \in E(S, V \setminus S)} w (e) }{ \min \set{ {\sf vol}(S), {\sf vol}(V \setminus S) }} \mcom\]
where $E(S, V \setminus S) \defeq \set{ \set{i,j} \in E : i \in S, j \notin S }$ and
${\sf vol}(S) \defeq \sum_{i \in S} \sum_{j \sim i} w \paren{\set{i,j}}$. The edge expansion of the graph is defined as
$\phi_G \defeq \min_{S \subset V, \ S \neq \emptyset} \phi_G(S)$.
\end{definition}
The {\em Stochastic Block Model} (we will refer to it as the {\em edge expansion stochastic block model} to
differentiate it from our block model) is a randomized model for instances that are generated as follows. 
A set of $n$ vertices is arbitrarily partitioned into sets $S,S'$ of equal sizes. Between each pair of 
vertices in $S$, an edge is added independently with probability $p$, and between each pair of vertices in
$S \times S'$, an edge is added independently with probability $q$ (typically $p > q$).

Starting with work of Holland \etal \cite{hll83},
the works of Boppana \cite{b87}, who gave a spectral algorithm, 
and of Jerrum and Sorkin \cite{js98}, who gave a metropolis algorithm, contributed significantly
to the study of stochastic block models. 
One of the break through works in the study of SBMs
is the work of McSherry \cite{m01}, who gave a simple spectral algorithm for a 
certain range of parameters.
There has been a lot of recent work related to a certain conjecture regarding SBMs,
which stated the regime of parameters $p,q$ for which it was possible to detect 
the presence of communities.
Works due to \cite{mns14a,mns15a,mns15b,m14} have contributed to proving various aspects
of the conjecture.
In a recent work, Abbe \etal \cite{abh16} showed that the natural \sdp~relaxation for 
balanced edge expansion is integral when there is a sufficient gap between $p$ and $q$, and
$p,q = \Omega(\log n/n)$; Mossel \etal \cite{mns15b} gave an algorithm for a 
larger regime of parameters which was not based on semidefinite programming.
More general SBMs have been studied by Abbe and Sandon \cite{as15a,as15b,as17},
Aggarwal \etal \cite{abkk15}, etc.

Kim \etal \cite{kbg17} studied a version of SBM for hypergraphs, and gave algorithms for it based
on studying a certain ``adjacency tensor'', the analog of the adjacency matrix for hypergraphs.
They also study the sum-of-squares algorithms for this model.
\cite{lm14b} gave a reduction from vertex expansion problems to hypergraph expansion problems. 
We note that applying this reduction to the instances from our models does not give the model 
studied by \cite{kbg17}: this reduction will only introduce hyperedges between the sets corresponding
to $T$ and $T'$, whereas the model studied by \cite{kbg17} adds random hyperedges between 
$S$ and $S'$. Moreover, many parts of a graph from our model are adversarially chosen.

\paragraph{Semi-random models for edge expansion problems.}
Monotone adversarial errors in SBMs are the arbitrary addition of edges between
pairs of vertices within $S$ (resp. $S'$), and the arbitrary deletion of existing 
edges between $S$ and $S'$. 
Feige and Kilian \cite{fk01} gave an algorithm for the edge expansion model with 
monotone adversarial errors when the gap between $p$ and $q$ is sufficiently large.
Guedon and Vershynin \cite{gv16} gave an algorithm based on semidefinite programming 
for partially recovering the communities for certain ranges of parameters. 
Moitra \etal \cite{mpw16} gave algorithms (based on semidefinite programming) 
and lower bounds for partial recovery in the 
stochastic block model with a monotone adversary. 
Makarychev \etal \cite{mmv16} gave an algorithm for partial recovery
for the stochastic block model with a monotone adversarial errors and a small
number of arbitrary errors (i.e. non-monotone errors). 

Makarychev \etal \cite{mmv12,mmv14} studied some semi-random models of instances for 
edge expansion problems.
In particular, \cite{mmv12} studied a model analogous to \vbmparamsp for edge expansion problems;
they showed that if the number of edges crossing $(S,S')$ is $\e m$, and if there is a set of
$m$ edges $E_1$ such that $(S, E_1)$ is a regular graph having spectral expansion at least $\Omega(\e)$,
then there is an algorithm to recover a balanced cut of edge expansion $\bigo{\e}$.
The proof of \prettyref{thm:vbmp} and that of the corresponding result in \cite{mmv12} 
both proceed by using the expansion of the underlying subgraph to show that an
$\Omega(n)$ sized subset of the SDP vectors lie in a ball of small radius. 
\cite{mmv12} use this to recover a constant factor bi-criteria approximation
to balanced edge expansion; we adapt this approach to vertex expansion to 
prove \prettyref{thm:vbmp}.

The results cited here are only a small sample of the work on the SBMs. 
Since our model is very different from the edge expansion stochastic block models,
we only give a brief survey of the literature here, and we refer the reader to
a survey by Abbe \cite{a17} for a comprehensive discussion.
In general, algorithms for edge expansion problems can not be used for our \vbmfull~
since sparse edge cuts and sparse vertex cuts can be uncorrelated; we give an example to 
illustrate this fact in \prettyref{app:edgeexpansionbad}.  In particular, the action of the monotone adversary in \vbm~rules out the use of edge-expansion based algorithms for detecting $S$ and $S'$.

\paragraph{Vertex Expansion.}

There has been some work in investigating vertex expansion (balanced and non-balanced) the worst-case setting. Bobkov \etal~\cite{bht00} gave a Cheeger-type inequality for vertex expansion, where a parameter $\lambda_\infty$ plays a role analogous to the use of the second eigenvalue $\lambda_2$ in the edge-expansion variant.
Feige \etal \cite{fhl08} gave a $\bigo{\sqrt{\log n}}$-approximation algorithm
for the problem of computing the vertex expansion of graphs.
Louis \etal \cite{lrv13} gave an \sdp~rounding based algorithm that computes a set having
vertex expansion at most $\bigo{\sqrt{\phiv_G \log d}}$, where $d$ is the maximum vertex degree;
they also showed a matching hardness result based on the Small-set expansion hypothesis.
Louis and Makarychev~\cite{lm14b} gave a bi-criteria approximation for Small-set vertex expansion, a problem related to vertex expansion. Chan \etal~\cite{cltz18} studied various parameters related to hypergraphs, including parameters related to hypergraph expansion; they showed that many of their results extend to the corresponding vertex expansion analogues on graphs.

\cite{lr14} studied a model of instances for vertex expansion similar to ours. In their
model, the adversary partitions the vertex set $V$ into two equal sized sets $S,S'$,
and chooses a subset $T$ (resp. $T'$) of $S$ (resp. $S'$) of size at most $\e n$. 
Next, the adversary chooses an arbitrary subset of pairs of vertices in $S$ (resp. $S'$)
to form edges such that graph induced on $S$ (resp. $S'$) is an edge expander. 
The adversary chooses an arbitrary subset of the pairs of vertices in $T \times T'$
to form edges. 
\cite{lr14} give an \sdp~rounding based algorithm to compute a set having vertex expansion
$\bigo{\sqrt{\e}}$; we reproduce their proof in \prettyref{app:lr14}.

\subsection{SDP Relaxation}
\label{sec:sdp}
We use the \sdp~relaxation for $\phivb_G$ (\prettyref{sdp:primal}),
this \sdp~is very similar to that of \cite{lrv13}. 
We give the dual of this \sdp~in \prettyref{sdp:dual}
(we show how to compute the dual \sdp~in \prettyref{app:dual}).

\begin{boxedminipage}{.5\linewidth}
\begin{SDP}[Primal]
\label{sdp:primal}
	\[ \min \sum_{i \in V} \eta_i \]
\subjectto
\begin{align*}
	U_{ii} + U_{jj} - 2 U_{ij} & \leq \eta_i & \forall i \in V, j \in N(i) \\
	U_{ii} & = 1 & \forall i \in V \\
	\sum_{i \in V} \sum_{j \in V} U_{ij} & = 0 \\
	U & \succeq 0
\end{align*}
\end{SDP}
\end{boxedminipage}%
\begin{boxedminipage}{.5\linewidth}
\begin{SDP}[Dual]
\label{sdp:dual}
\[ \max \sum_{i \in V} B_{ii} \]
\subjectto
\begin{align*}
	\sum_{j \in N(i)} Y_{ij} & = 1 \\
	Y_{ij} & \geq 0		& \forall \set{i,j} \in E \\
	B_{ij} & = 0 		& \forall i,j \in V, i \neq j \\ 
	L(Y) + \alpha \one \one^T - B & \succeq 0
\end{align*}
\end{SDP}
\end{boxedminipage}

Here $\one$ denotes the all-ones vector, and $L(Y)$ denotes the Laplacian matrix of graph weight by the matrix $Y + Y^T$, i.e.
\[ L(Y)_{ij} = \begin{cases}	\sum_{l \in N(i)} \paren{Y_{il} + Y_{li}} & i = j \\
				-\paren{Y_{ij} + Y_{ji}}		& j \in N(i) \\
				0	& \textrm{otherwise}	\end{cases} \mper   \]
First, let us see why \prettyref{sdp:primal} is a relaxation for $\phivb$.
Let $P$ be the set corresponding to $\phivb_G$, and let 
$\one_P \in \set{-1,1}^n$ be a vector such $\one_P(i)$ is equal to $1$ if 
$i \in P$ and $-1$ otherwise. 
Note that since $\Abs{P} = \Abs{V}/2$, we have $\one^T \one_P = 0$. 
It is easy to verify that 
$U := \one_P \one_P^T$ and $\eta_i := \max_{j \in N(i)} \paren{\one_P(i) - \one_P(j)}^2 $
is a feasible solution for \prettyref{sdp:primal}, and that 
$\sum_{i \in V} \eta_i = 4\paren{\Abs{N(P)} + \Abs{N(V \setminus P)}}$. 
Therefore, $\phivb_G = 4 \paren{\sum_{i \in V} \eta_i}/n$.  
and 
therefore, \prettyref{sdp:primal} is a relaxation for $\phivb$.
Henceforth, we will use $\one_P$ to be the indicator vector of a set $P \subset V$, i.e.,
$\one_P(i)$ is equal to $1$ if $i \in P$ and $-1$ otherwise.
We prove the following theorem about \prettyref{sdp:primal}.
\begin{theorem}
\label{thm:sdpintegral}
For the regime of parameters stated in \prettyref{thm:main}, 
$U' \defeq \inds \inds^T$ and for each $i$, $\eta_i' \defeq \max_{j \in N(i)} \paren{\inds(i) - \inds(j)}^2$ 
for the set $S$ defined in \vbmparams, 	is the unique optimal solution to \prettyref{sdp:primal}
with probabilty at least $1 - 1/{\sf poly}(n)$.
\end{theorem}
\prettyref{thm:sdpintegral} gives an algorithm to compute the matrix $\inds \inds^T$. By factorizing this
matrix, one can obtain the vector $\inds$, using which the set $S$ can be computed. 
Therefore, \prettyref{thm:sdpintegral} implies \prettyref{thm:main}.

In \prettyref{sec:const-factor-approx}, we give a rounding algorithm for \prettyref{sdp:primal},
which we use to prove \prettyref{thm:vbmp} and \prettyref{thm:vbmlambda}.

\subsection{Proof Overview}
\label{sec:proofoverview}
\subsubsection{\prettyref{thm:main}}
It is easy to verify that $\paren{U',\eta_i'}$  is a feasible solution to \prettyref{sdp:primal}.
Our goal will be to construct a dual solution (i.e. a feasible solution to
\prettyref{sdp:dual}) which satisfies two properties, 
\begin{enumerate}
	\item The cost of this solution should be same as the cost 
		of this primal solution $(U',\eta')$.
	\item The matrix $\paren{L(Y) + \alpha \one \one^T - B}$ should have rank $n-1$.
\end{enumerate}

Using strong duality, (1) will suffice to ensure that $(U',\eta')$ is an optimal 
solution of the primal \sdp. 
To show that this is the unique primal optimal solution,
we will use the complementary slackness conditions which state that 
\begin{equation}
\label{eq:cs}
	U \cdot \paren{L(Y) + \alpha \one \one^T - B} = 0 \mper
\end{equation}
For the sake of completeness, we give a proof of this in \prettyref{app:dual}.
Since, $\paren{L(Y) + \alpha \one \one^T - B}$ will have rank $n-1$, this will imply that 
all primal optimal solutions must have rank at most $1$, or in other words,
there is a unique primal optimal solution (see \prettyref{lem:primal-integ}). 

While the approach of using complementary slackness conditions for proving the integrality of the SDP relaxation has been studied for similar problems before (\cite{coja2007, abbs2014,abh16,hwx2016,abkk2017}), there is no known generic way of implementing this approach to any given problem. Usually the challenging part in implementing this approach is in constructing an appropriate dual solution, and that, like in most of the works cited above, forms the core of our proof.

We give an outline of how we construct our dual solutions. 
We begin by setting the $Y$ value for each edge added by the monotone adversary to $0$,
thus our proof can be viewed as saying that \prettyref{sdp:primal} ``ignores'' all those edges. 
For the sake of simplicity, let us consider the case when the bipartite graph between 
$T$ and $T'$ is a $c$-regular graph. We set $B_{ii} := 4$ if $i \in T \cup T'$ and $0$
if $i \notin T \cup T'$.
Thus, if we can choose $Y$ such that this choice of $B$ is a feasible solution, then
this will ensure that the cost of this dual solution, and the cost of the primal solution $(U',\eta')$ are
both equal to $4 \paren{\e_1 + \e_2} n$, thereby fullfilling our first requirement.

If $U$ is a rank one matrix, and $\paren{L(Y) + \alpha \one \one^T - B}$ is a rank $n-1$ matrix, 
then \prettyref{eq:cs} implies that $\inds$ is an eigenvector of $\paren{L(Y) + \alpha \one \one^T - B}$
with eigenvalue $0$. 
This fact will be extremely useful in setting the $Y$ values 
for the edges in the bipartite graph between $T$ and $T'$
(\prettyref{lem:setting-B}).
Now, we only have to choose the $Y$ values for the edges fully contained in $S$
(resp. $S'$).
We first prove the following lemma which will help us to choose the $Y$ values.
\begin{lemma}[Informal statement of \prettyref{lem:suff-cond-lemma}]
\label{lem:suff-cond-informal}	
There exists a constant $c'$ such that it suffices to choose $Y$ satisfying 
\[ X^T \paren{L(Y)} X \geq  \frac{c'}{n} \paren{ \sum_{i \in S \setminus T} \sum_{t \in T} 
		\paren{X_i - X_t}^2 + \sum_{i \in S' \setminus T'} \sum_{t \in T'} 
		\paren{X_i - X_t}^2}  \qquad \forall X \in \R^n \mper \]
\end{lemma}
The proof of this lemma follows by carefully choosing the value of $\alpha$, and by 
exploiting the fact that $\inds$ is an eigenvector of $L(Y) + \alpha \one \one^T - B$
with eigenvalue $0$.
Proving the condition in \prettyref{lem:suff-cond-informal} can be viewed as the problem of 
choosing capacities for the edges to support the multicommodity flow where
each vertex $i \in S \setminus T$ wants to send $c'/n$ amount of flow to each $t \in T$.  
This idea can work when $S$ (resp. $S'$) is a sufficiently dense graph, but does not work when 
$S$ (resp. $S'$) is sparse (\prettyref{rem:flow-lower-bound}). 
Our second idea is to use the edge expansion properties of the underlying spanning subgraph. 
For a $d$-regular edge expander $H = (V',E')$ having the second smallest normalized Laplacian 
eigenvalue $\lambda$, we get that 
$\sum_{\set{ij} \in E'} \paren{X_i - X_j}^2 \geq \paren{\lambda d/n}\sum_{ij \in E'} \paren{X_i - X_j}^2$.
Since $L(Y)$ is a Laplacian matrix, we get that 
\[ X^T L(Y) X = \sum_{\substack{i,j \in S \\ \set{i,j} \in E}} \paren{Y_{ij} + Y_{ji}} \paren{X_i - X_j}^2 
	+ \sum_{\substack{i,j \in S' \\ \set{i,j} \in E}} \paren{Y_{ij} + Y_{ji}} \paren{X_i - X_j}^2
	+ \sum_{\substack{i \in T, j \in T' \\ \set{i,j} \in E}} \paren{Y_{ij} + Y_{ji}} \paren{X_i - X_j}^2  \mper \]
Now, since $S$ and $S'$ contain an almost regular edge expander as a spanning subgraph, we can adapt the expander
argument to this setting and obtain some lower bound on this quantity. 
This strategy can work in some special cases, but fails in general
(\prettyref{rem:expansion-lower-bound}).
Our proof shows that the desired lower bound in \prettyref{lem:suff-cond-informal} can be obtained using a
careful combination of these two ideas, in addition to exploiting the various properties of the random graph
between $S \setminus T$ and $T$ (resp. $S' \setminus T'$ and $T'$).

\subsubsection{\prettyref{thm:vbmlambda} and \prettyref{thm:vbmp}}
We first solve \prettyref{sdp:primal} and obtain a matrix $U$ such that $U \succeq 0$. Therefore, $U$ can be
factorized into $U = W^T W$ for some matrix $W$. Let $u_1, \ldots, u_n$ denote the columns of this matrix $W$. 
We give an algorithm (see \prettyref{sec:const-factor-approx}) to ``round'' these vectors into a set satisfying the guarantees in the 
theorem.
As in the previous case, we show that we can ``ignore'' all the edges added by the
monotone adversary, and only focus on the edges added in \prettyref{step:rand}, \ref{step:arbs}, \ref{step:tt} 
in \prettyref{def:vbm}. 

A well known fact for edge expander graphs having roughly equal vertex degrees is that if the value of 
$\norm{u_i - u_j}^2$ averaged over all edges $\set{i,j}$ in the graph is small, then the value of 
$\norm{u_i - u_j}^2$ averaged over all pairs of vertices $i,j$ in the graph is also small. 
In the proof of \prettyref{thm:vbmp}, we use the expansion properties of the $\Omega(n)$ sized subset of $S$ coupled with this fact
to show that an $\Omega(n)$ sized subset of the vectors $\set{u_i : i \in V}$ must 
lie in a ball of small diameter; this step is similar to the corresponding step of \cite{mmv12}.
We use this to construct an embedding of the 
graph onto a line, and recover a cut from this embedding 
using an algorithm of \cite{lrv13}; this step can be viewed as adapting the corresponding step of 
\cite{mmv12} to vertex expansion.

In the case when $\lambda_1 = 0$, 
we show that the lopsided random bipartite graph between $S \setminus T$ and $T$ 
is an edge expander w.h.p. However, this graph is not close to being regular;
the degrees of the vertices in $T$ would be much higher than the degrees of the 
verticies in $S \setminus T$. Therefore, we can not directly use the strategy employed in the
previous case. But we show that we can use the fact that the measure of $S \setminus T$ under the
stationary distribution of the random bipartite graph between $S \setminus T$ and $T$ is $\Omega(1)$,
and that the vertices in $S \setminus T$ have roughly equal vertex degrees, to show that 
$\norm{u_i - u_j}^2$ averaged over all pairs of vertices $i,j \in S \setminus T$ is small. 
From here, we proceed as in the previous case.

\newrefformat{obs}{\savehyperref{#1}{Observation~\ref*{#1}}}
\newrefformat{fct}{\savehyperref{#1}{Fact~\ref*{#1}}}
\newcommand{\hatG}{\widehat{G}}
\newcommand{\hatY}{\widehat{Y}}
\newcommand{\dout}{d^{\mathrm{out}}} 

\subsection{Notation}\label{sec:notation}
We denote graphs by $G=(V,E)$, where the vertex set $V$ is identified with $[n]\eqdef \inbrace{1, 2, \ldots n}$. The vertices are indexed by $i,j$, or, if belonging to the specific subset $T \subseteq V$ ($T' \subseteq V$) in the \vbm~model, we use $t$ (resp. $t'$) for clarity. The optimal value of the vertex expansion on an instance $G$ is denoted by $\phiopt$, and the value attained by the algorithm is denoted by $\phialg$. The value of the primal SDP relaxation for vertex expansion on $G$ is denoted by $\phisdp(G)$, and the value of the dual by $\phidual(G)$. For any $S \subseteq V$, we denote the induced subgraph on $S$ by $G[S]$. Given $i \in V$ and $T \subseteq V$, define $N_T(i) \defeq \set{j \in T ~:~ \set{i,j} \in E}$, and $N(i) = N_V(i)$. We denote $\Delta_T(i) \eqdef |N_T(i)|$, and $\Delta(i) = |N(i)|$. For a subgraph $\mF$ of $G$, the degree of $i$ within $\mF$ will correspondingly be $\Delta_\mF (i)$.

\medskip
Given a graph $G=(V,E)$ with a weight $w: E \rightarrow \R_{\geq 0}$ on its edges, we define the \emph{weighted} degree of a vertex $i\in V$ as $d(i) \eqdef \sum_{j \in N(i)} w_{\inbrace{i,j}}$. The (un-normalized) Laplacian $L \in \R^{n\times n}$ of a graph $G=(V,E)$ with a weight function $w_{\inbrace{i,j}}$ on its edges is given by $L=D-A$, where $ D_{ij} = \diag(d(1), \ldots, d(n))$ and $A_{ij} = w_{\inbrace{i,j}}$. Similar to the unweighted degrees, for any $S \subseteq V$, we define $d_S(i) \defeq \sum_{j \in N(i)\cap T} w_{\inbrace{i,j}}$. We will call a graph as \emph{close-to-regular} or \emph{almost regular}, if the ratio $\frac{\max_i \Delta(i)}{\min_i \Delta(i)}$ is at most some constant.

\medskip
Typically, for a vector $X \in \R^n$, its $i$-th component is denoted by $X_i$, or in rare cases for clarity, by $X(i)$. The Hadamard product of two matrices $A, B \in \R^{n \times n}$  is denoted by $A \cdot B \defeq \trace (A^TB) = \sum_{i,j} A_{ij}B_{ij}$.
As an exception, when we are dealing with vectors associated by the SDP solutions to the vertices of a graph, we exclusively use $u_i \in \R^n$ to be the vector associated with vertex $i$.

\medskip
We note that for any vector $X\in \R^n$, we have $X^T LX = \sum_{\set{i,j} \in E} w_{\inbrace{i,j}} (X_i - X_j)^2$. For a $S\subseteq V$,  we denote $X^T L_{\mid S}X = \sum_{\set{i,j} \in E(G[S])} w_{\inbrace{i,j}} (X_i - X_j)^2$

\medskip
In our proofs, following \prettyref{sdp:dual}, we will be assigning directed weights (or capacities) $Y_{ij}$ to edges $\set{i,j}\in E$, and use  $L(Y)$ to denote the  Laplacian of the graph with weights $Y_{ij}+Y_{ji}$ on the edges. Often, when clear from context, we drop the argument $Y$ for clarity.

\medskip
Probability distributions $\mu$ will defined over some finite set $\Omega$. Given a random variable $X: \Omega \rightarrow \R$, its expectation is denoted by $\E_{\omega \sim \mu}[X]$.  When the distribution is not specified explicitly, it is assumed to be the uniform distribution on $\Omega$, and expectations with respect to the uniform distribution are written as $\E_{\omega \in \Omega}[X]$.

\medskip
We say that an event $E$ related to some graph $G$ occurs \emph{with high probability}, if $\Pr{E} \geq 1- {1}/{\poly(n)}$, where $n$ is the number of vertices in $G$.

\medskip
 Given an undirected graph $G=(V,E)$, denote the \emph{stationary distribution}  over the vertices by $\mu_G$, defined as $\mu_G(i) = \frac{\Delta(i)}{\sum_{j \in V}\Delta(j)}$. Given the normalized Laplacian $\mathcal{L} = I- D^{-1/2} A D^{-1/2}$, the \emph{spectral gap} of $G$ denoted by  $\lambda$, is the second-smallest eigenvalue of $\mathcal{L}$. \emph{Spectral expanders} are a family of graphs with $\lambda$ at least some constant (independent of the number of vertices in $G$).

\medskip
As in the introduction, we use $\one_S$ for any $S \subseteq V$ to denote the vector in $\R^{|V|}$ having entries $\one_S(i) = 1$, if $i \in S$, and $-1$ otherwise.

\section{Exact Recovery for \vbm}\label{sec:proof}

\subsection{A sufficient condition}
In order to prove \prettyref{thm:sdpintegral}, we first prove the following lemma, which outlines a sufficient condition for integrality of the primal optimal \sdp~solution.

\begin{lemma}\label{lem:suff-cond-lemma}
 For a \vbmparams~  instance, if there exists a $Y \in \R^{n\times n}$ that satisfies:  
 
\begin{enumerate}
\item[(a)] $\forall i,j~:~ ~Y_{ij} \geq 0$ , 
\item[(b)] $\forall i \in V~:~ \sum_{j\in N(i)}Y_{ij} =1$, 
\item[(c)] \begin{align}
X^T \paren{L(Y)} X \geq  \frac{c'}{n} \paren{ \sum_{i \in S \setminus T} \sum_{t \in T} 
		\paren{X_i - X_t}^2 + \sum_{i \in S' \setminus T'} \sum_{t \in T'} 
		\paren{X_i - X_t}^2} \qquad \forall X \in \R^n \mcom \label{eq:suff-cond}
\end{align}
\item[(d)] For every $i \in T, j \in T'$ and $\{i,j\} \in \mF$, we have $Y_{ij} = \frac{1}{\Delta_{\mF}(i)}$ and $Y_{ji} = \frac{1}{\Delta_{\mF}(j)}$,
\end{enumerate}
where $c' = 8c/(1-\max \set{\e_1, \e_2})$, then \prettyref{sdp:primal} has $(U', \eta')$ as defined in \prettyref{thm:sdpintegral} as its unique optimal solution.
\end{lemma}

\begin{remark}
While conditions (a) and (b) are explicitly part of the \prettyref{sdp:dual} constraints, the remaining conditions (c) and (d) together ensure that we can extend $Y$ to a feasible dual solution $(Y, B, \alpha)$, that satisfies the positive-semidefiniteness constraint \emph{and}  is optimal. As $\mF$ is a spanning forest on the bipartite subgraph on $T \cup T'$, the weights mentioned in condition (d) are always well-defined. 
\end{remark}

\medskip

\begin{proof}[Proof (of ~\prettyref{lem:suff-cond-lemma})]
We begin by noting a simple consequence of the complementary slackness conditions. We drop the argument $Y$ from $L(Y)$ as it is clear from context.

\begin{lemma}\label{lem:primal-integ}
Let $M \defeq L+\alpha \one \one^T - B$ be constructed using an optimal dual solution $(Y,B, \alpha)$.  The primal optimal solution is integral and unique if $\one_S$ is a unique eigenvector of $M$ with eigenvalue $0$. 
\end{lemma}

\begin{proof}
Suppose one of the optimal solutions of rank $k>1$. Consider the spectral decomposition of $U$ 
\[
U = \sum_{i \in [k]} \lambda_i v_i v_i^T 
\]
where $\lambda_i > 0$ are the eigenvalues, and $v_i$ are the eigenvectors of $U$.  Since complementary slackness (See \prettyref{app:dual}) implies that $U \cdot M = 0$ , we get $\sum_{i \in [k] } v_i^T M v_i =0$. Since $M \succeq 0$, we should have $v_i^T M v_i =0$, for each $i \in [k]$, meaning every $v_i$ is a zero eigenvector of $M$. This is a contradiction for $k>1$, since $M$ has a unique $0$ eigenvector.
Thus, $U$ is rank $1$, and by the assumption, it is a linear multiple of $\one_S \one_S^T$. By the constraints in \prettyref{sdp:primal} saying $U_{ii}=1$, we get that $U = \one_S \one_S^T$.
\end{proof}

It is thus sufficient to prove that the conditions in \prettyref{lem:suff-cond-lemma} imply that we can use the given $Y$ to come up with a $B$ and $\alpha$, such that $(Y,B, \alpha)$ is feasible, \emph{and} $\one_S$ is a unique eigenvector of $M$ with eigenvalue $0$. We first find a $B$ (depending on $Y$) that yields a dual objective  value of exactly $4(\e_1 + \e_2) n$,  and ensures that $\one_S$ is an eigenvector with eigenvalue $0$. Recall that $d_T(i) = \sum_{j \in N(i)\cap T} (Y_{ij} + Y_{ji})$ is the weighted degree of $i$ into $T \subseteq V$.

\begin{observation} \label{obs:weights-yts-zero}
For every $i\in T, j \in S$ and $i \in T', j \in S'$, we have $Y_{ij}=0$. Further, $Y_{ij}=Y_{ji}=0$ if $i \in T, j \in T'$ and $\set{i,j} \notin \mF$.
\end{observation}
\begin{proof}
Consider an $i \in T$. Condition (d) in \prettyref{lem:suff-cond-lemma} already gives us that $
\sum_{j \in N_{\mF}(i)} Y_{ij}=1$. For any other $j \in N(i) \setminus N_{\mF}(i)$, we should have $Y_{ij}=0$, as $Y_{ij} \geq 0$ are all non-negative, and $\sum_{j \in N(i)} Y_{ij}=1$. A similar argument holds, if $i \in T'$. 
\end{proof}

\begin{lemma}\label{lem:setting-B}
Fix some partial candidate dual solution $(Y,\alpha)$. Consider the diagonal matrix $B$ given by:
\begin{align}
 B_{ii} = \left\{\begin{array}{lr}
        2\cdot d_{T'}(i), & \text{for } i\in T\\
        2\cdot d_{T}(i), & \text{for } i\in T'                    \\        
        0, & \text{otherwise }
        \end{array}\right . 
  \end{align}
  
Then $\one_S$ is an eigenvector of $M$ with eigenvalue $0$. Furthermore, if $(Y, \alpha)$ is feasible for this $B$ and satisfies:

\begin{align}\label{eq:cond-on-y}
Y_{ij} = 0 \mcom \quad \text{if } i\in T, \, j\in S \text{ or } i\in T',\, j\in S'  \mcom
\end{align}
then the dual variable assignment $(Y,B, \alpha)$ is optimal, with objective value $4(\e_1 + \e_2) n$.
\end{lemma}

\begin{proof}
To prove the first part, we show that $M \one_S = 0$. To see this, fix some $i \in T$ (a similar argument holds for $i \in T'$), and consider that:
\begin{align*}
(L + \alpha \one \one^T) \one_S  ~&=~  L\one_S \\
\implies \paren{(L + \alpha \one \one^T) \one_S}_i ~&=~ d_{T'}(i)\one_S(i) - \sum_{j \in N_{T'}(i)} \one_S(j) (Y_{ij}+Y_{ji})\\
~&=~ 2 \cdot d_{T'}(i)\\
~&=~ B_{ii} 
\end{align*}

The first equality follows from the fact that $\one_S \perp \one$. The second equality is due to the fact that within $S$ or $S'$, $\one_S$ is a constant, and hence edges within these do not contribute to the sum. Thus, we only need to look at edges of $i$ across the bipartite graph on $T \times T'$. The final step used the fact that since we are within the bipartite subgraph on $T\times T'$, we have $\one_S(i) = - \one_S(j)$ for $\set{i,j} \in E(G)$. The above implies that $(L+\alpha \one \one^T)\g = B\g ~\implies~ (L+\alpha \one \one^T- B)\g=0$. By the definition of $M$, we infer that it has $\one_S$ as an eigenvector.

\medskip

In order to prove the second part of the lemma, let $(Y,\alpha)$ be a feasible solution pair for the above $B$ that satisfies the given conditions.  Then, we have that the dual objective value is:

\begin{align*}
\sum_{i \in V} B_{ii} ~&=~ 2 \cdot \sum_{i \in T} d_{T'}(i) + 2 \cdot \sum_{i \in T'} d_{T}(i) 
~\stackrel{(a)}{=}~ 2 \sum_{\substack{i\in T\cup T'\\ j \in N_{T\cup T'}(i)}} (Y_{ij} + Y_{ji})\\
~&\stackrel{(b)}{=}~ 4 \sum_{\substack{i\in T\cup T'\\ j \in N_{T\cup T'}(i)}} Y_{ij}
 ~\stackrel{(c)}{=}~ 4 \times (|T| + |T'|)  = 4 (\e_1 + \e_2) n\mper  
\end{align*}

Above, (a) follows from the definition of $d_{T}$ and $d_{T'}$, (b) follows from the fact that every such $Y_{ij}$ appears exactly twice in the previous sum. Finally, $(c)$ follows from the fact that the weights $Y_{ij}$ from $i \in T$ to any $j \in S\setminus T$ are $0$ from \prettyref{obs:weights-yts-zero}, and the dual \sdp (\prettyref{sdp:dual}) sets the sum of the weights $Y_{ij}$ out of every node $i$ to be equal to $1$. Since $\mF$ is a spanning subgraph, every node in $T \cup T'$ contributes exactly $1$ to the sum. 

\medskip
Since the primal has a feasible integral solution of value $4(\e_1+ \e_2) n$,  it follows that such a feasible dual solution  $(Y,B, \alpha )$ is indeed optimal.
\end{proof}

\medskip

As \prettyref{obs:weights-yts-zero} shows that the conditions in \prettyref{lem:suff-cond-lemma} cover the conditions required on a candidate $Y$ in \prettyref{lem:setting-B}, it now remains to show that a $Y$ that obeys the preconditions in \prettyref{lem:suff-cond-lemma} satisfies:

\[
X^T \paren{L(Y) + \alpha \one \one^T} X ~\geq~ X^T BX  \qquad \forall X \in \R^n \mper
\] 

\medskip
We first simplify the RHS in the above equation. Condition $(d)$ in \prettyref{lem:suff-cond-lemma} sets $Y_{ij} = \frac{1}{\Delta_{\mF}(i)}$ for every edge between $i \in T$ and $j \in T'$ in $\mF$ (and similarly for $Y_{ji}$).  We use the setting for $B$ specified by \prettyref{lem:setting-B}. This gives us that for every $i\in T\cup T'$, $B(i,i) \leq 2c$, since every edge incident on $i$ can have weight at most $2$. Thus, we have:  

\begin{equation} \label{eq:bound-on-xBx}
X^T BX ~\leq~ 2c ~\sum_{t \in T} X_t^2.
\end{equation}

To tackle the LHS, we will use the following fact:

\begin{fact}\label{fct:matrix-trick}
If $M\in \R^{n \times n }$ is a symmetric matrix with eigenvector $v$ having eigenvalue $0$, then:
\begin{equation}\label{eq:ggT-trick}
M \succeq 0  \text{ and } \mathrm{rank} (M)=n-1 \iff \exists l > 0 ~:~ M ~+~ l\cdot v v^T \succ 0
\end{equation}
\end{fact}
\begin{proof}
The forward implication is straightforward, and it in fact holds for all $l > 0$. For the reverse implication:  if $M$ has a negative eigenvalue, then clearly, adding $ l v v^T$ changes only the eigenvalue corresponding to $v$.   Thus, $M+ l\cdot vv^T$ will continue to have a negative eigenvalue. 
\end{proof}

We will use this fact with $M = (L+ \alpha \one \one^T -B)$, and $v = \one_S$; by our setting for $B$, $\one_S$ is an eigenvector of $M$ with eigenvalue $0$. We first state and prove some lemmas which we use to prove \prettyref{lem:suff-cond-lemma}.

\begin{lemma}\label{lem:helper-lemma-1}
Let $\abs{T} = \eps n$. For any $X \in \R^n$, and $b,l \geq 0$, we have:
\[
2b~\paren{\sum_{t \in T} X_t}\paren{\sum_{i \in S \setminus T } X_i} ~-~ 2l~ \paren{ \sum_{i \in S} X_i}^2 \leq \frac{b^2}{2l}\times \frac{\epsilon n}{2} \sum_{t \in T} X_t^2   
\]
\end{lemma}

\begin{proof}
We have:  $$2b~\paren{\sum_{t \in T} X_t}\paren{\sum_{i \in S \setminus T } X_i} = 2b~\paren{\sum_{t \in T} X_t}\paren{\sum_{i \in S} X_i} - 2b \paren{\sum_{t \in T} X_t}^2 \mper $$  Substituting into the LHS gives us:

\begin{align*}
\mathrm{LHS}  &~=~ 2b~\paren{\sum_{t \in T} X_t}\paren{\sum_{i \in S} X_i} - 2b \paren{\sum_{t \in T} X_t}^2 ~-~ 2l~ \paren{ \sum_{i \in S} X_i}^2 \\
& \leq \frac{b^2}{2l} \paren{\sum_{t \in T} X_t}^2 ~\leq~ \frac{b^2}{2l}\times \frac{\epsilon n}{2} \sum_{t \in T} X_t^2\\
\end{align*}

The first inequality holds because for any $\alpha, \beta \in \R$, we have  $$2b \alpha \beta - 2b \alpha^2 - 2l \beta^2 = \frac{ b^2} {2l} \alpha^2 -2b \alpha^2 - \paren{\frac{b}{\sqrt{2l}}\alpha - \sqrt{2l} \beta }^2 \mcom$$ where $b, l \geq 0$, and the second step follows by an application of the Cauchy-Schwarz inequality.
\end{proof}

\medskip

\begin{lemma}\label{lem:helper-lemma-2} Let $|T| = \eps n$. For any $b,l \geq 0$, and $X \in \R^n$,
\begin{align}
b \sum_{\substack{i\in S\setminus T \\ t \in T}} (X_i - X_t)^2 ~+~ 2l \paren{\sum_{i\in S} X_i}^2 &\geq  \paren{\frac{(1-\epsilon) bn}{2} - \frac{\epsilon b^2 n}{4l}} \sum_{t \in T} X_t^2   +b \sum_{i \in S\setminus T} X_i^2
\end{align}

\end{lemma}
 \begin{proof}
This follows by expanding out the LHS terms:
\begin{align*}
& b\sum_{\substack{i\in S\setminus T \\ t\in T}} (X_i - X_t)^2 ~+~ 2l \paren{\sum_{i\in S} X_i}^2  \\
&= b \epsilon n\sum_{i \in S \setminus T} X_i^2 + \frac{(1-\epsilon)bn}{2} \sum_{t\in T} X_t^2 - 2b~\paren{\sum_{t \in T} X_t}\paren{\sum_{i \in S \setminus T } X_i} + 2l~ \paren{ \sum_{i \in S} X_i}^2\\
& \geq b \epsilon n \sum_{i \in S \setminus T} X_i^2 ~+~ \frac{(1-\epsilon)bn}{2} \sum_{t\in T} X_t^2 - \frac{b^2}{2l}\times \frac{\epsilon n}{2} \sum_{t \in T} X_t^2 \qquad \ldots \text{Using \prettyref{lem:helper-lemma-1}}\\
&= RHS.
\end{align*}
\end{proof}

 We are now ready to complete the proof of \prettyref{lem:suff-cond-lemma}. We are only left with determining that the precondition in the lemma yields the required results.  Note that: 

\begin{equation}\label{eq:ss'}	
X^T \paren{ \g \g ^T + \one \one^T} X = 2 \paren{\sum_{i \in S} X_i}^2 + 2 \paren{\sum_{i \in S'} X_i}^2 \mper
\end{equation}

The above equality holds because $\paren{\g \g ^T + \one \one^T}_{ij} = 1$ for $i,j \in S$ or $i,j \in S'$, and is $0$ otherwise. 

\medskip
In order to prove that $M \succeq 0$ and that it has $\mathrm{rank}(M)=n-1$ , in view of  \prettyref{fct:matrix-trick}, we instead show that there exists some $l> 0$ such that $M +   l \cdot \g \g^T \succ 0$, for some value of $l$ that we will choose later. Let us set $\alpha \seteq l$. Consider any $X \in \R^n$;  writing out explicitly the quantity $X^T (M +   l \cdot\g \g^T)X$ gives us:


\begin{align*}
X^T \paren{M + l \g \g^T} X ~&=~ X^T L X + l X^T \paren{\g \g^T + \one \one^T}X 
		- X^TBX \\
&\stackrel{(a)}{\geq} ~ X^TLX ~+~ l \cdot X^T \paren{\g \g^T + \one \one^T}X - 2c \cdot \paren{\sum_{t\in T} X_t ^2 + \sum_{t'\in T'} X_{t'} ^2} \\
&\stackrel{(b)}{\geq}  \frac{c'}{n} \paren{ \sum_{\substack{ i \in S \setminus T \\ t \in T}} \paren{X_i - X_t}^2 + \sum_{\substack {i \in S' \setminus T'\\ t' \in T'} } \paren{X_i - X_{t'}}^2}
	+2 l \paren{ \paren{\sum_{i \in S} X_i}^2 + \paren{\sum_{i \in S'} X_i}^2 } \\
	& \qquad \qquad- 2c \cdot \paren{\sum_{t \in T} X_t^2 + \sum_{t' \in T'} X_{t'}^2}\\	
	& \stackrel{(c)}{=}~ \frac{c'}{n} \sum_{\substack{ i \in S \setminus T \\ t \in T}} (X_i - X_t)^2 + 2 l  \paren{\sum_{i \in S} X_i}^2 - 2c \sum_{t \in T} X_t^2 \\
	&\qquad +~ \frac{c'}{n}  \sum_{\substack{ i \in S' \setminus T' \\ t' \in T'}} (X_i - X_{t'})^2 + 2l  \paren{\sum_{i \in S'} X_i}^2 - 2c \sum_{t' \in T'} X_{t'}^2 \\
	&\stackrel{(d)}{\geq}  \paren{\frac{(1-\epsilon) c'}{2} - \frac{\epsilon c'^2 }{4ln} - 2c} \sum_{t \in T \cup T'} X_t^2  + ~\frac{c'}{n}\sum_{i \notin (T\cup T')} X_i^2
	\end{align*}
	
Above, $(a)$ follows from \eqref{eq:bound-on-xBx}, $(b)$ follows from \eqref{eq:ss'} and the conditions in the statement of \prettyref{lem:suff-cond-lemma}, $(c)$ is a simple rearrangement of terms, and $(d)$ is obtained using \prettyref{lem:helper-lemma-2}, with $b \seteq c'/n$.

\smallskip

Since $l$ is not bounded above, we let $l \rightarrow \infty$. Since $c' = 8c/(1-\max \set{\e_1, \e_2})$, for this setting of parameters, we have that as long as $\exists i\in V: X_i \neq 0$, we have:  $M+ l\cdot \g \g^T\succ 0$, thus ensuring that $M$ has $\g$ as the only zero eigenvector (by \prettyref{fct:matrix-trick}). By \prettyref{lem:primal-integ}, we get that the primal \prettyref{sdp:primal} is integral and optimal, with $U'=\g\g^T$ as the unique optimal solution. This completes the proof of \prettyref{lem:suff-cond-lemma}.
\end{proof}

\subsection{Satisfying the sufficient condition}

Given \prettyref{lem:suff-cond-lemma}, we are now left with the task of showing the existence of appropriate (directed) weights $Y_{ij}$ to the edges. First, we recall that for any $X\in \R^n$:

\[
X^T L(Y) X = \sum_{\set{i,j}\in E} (Y_{ij} + Y_{ji}) (X_i - X_j)^2
\]

The following observation is useful to keep in mind:
\begin{observation}\label{obs:y-leq-1}
Given a $(Y, B, \alpha)$ dual solution, that satisfies all but the constraints $\sum_{j \in N(i)} Y_{ij}=1$, having instead that   $\forall i \in V~:~ \sum_{j \in N(i)} Y_{ij} \leq 1$, we can produce a feasible solution $(Y', B, \alpha)$ of the same objective value. 
\end{observation}
\begin{proof}
This follows by noting that since $Y_{ij}$'s are non-negative, the expression $X^T L(Y) X$ is monotonically non-decreasing with respect to each of the $Y_{ij}$'s. We can thus pick an arbitrary neighbor $j$ for every $i$, and set $Y'_{ij}= 1- \sum_{k \in N(i) : k \neq j} Y_{ij}$, and $Y'_{ij} = Y_{ij}$ for the rest of the neighbors. Since $Y' \geq Y$, we have that $(Y', B, \alpha)$ is now feasible, and has the same objective value.   
\end{proof}

We will henceforth find values for the dual variable $Y$ satisfying just the weaker constraint $\sum_{j \in N(i)} Y_{ij}\leq 1$. We also implicitly ignore any adversarial edge $\{i,j\}$ added in the last step of the \vbm~instance construction, by setting the corresponding $Y_{ij}, Y_{ji}$ values to zero. In our arguments below, the graph considered is the subgraph of $G$ without adversarial edges.

\medskip

Let us start by noting the following simple case, in order to aid intuition.

\begin{lemma}\label{lem:Kn-integrality}
When $G[S]$ and $G[S']$ are complete graphs, there is a constant $\beta <1$ and weights $Y$, such that when $\epsilon \defeq \max \set{\e_1, \e_2} \leq\beta$, the sufficient condition in \prettyref{lem:suff-cond-lemma} is satisfied, and hence the primal SDP is integral. 
\end{lemma}
\begin{proof}
For every pair $i,j$ such that $i\in S \setminus T$ and $t \in T$, we set $Y_{it} = b \seteq \frac{1}{\epsilon n}$. The $Y_{ij}$'s within $S'$ are set similarly. Let $Y_{ij}=0$ for all other edges within $G[S]$, and $G[S']$. The constraint $\sum_{j \in N(i)} Y_{ij}  \leq 1$ is satisfied as:

\begin{equation}\label{eq:outdeg-bound-Kn}
                b\times \Delta_T(i) \leq 1 \implies b \epsilon n \leq 1
\end{equation}

 In order to prove integrality, we verify that \eqref{eq:suff-cond} holds for the chosen value of $b$:

\begin{align*}
X^T L X ~&\geq~ b~\sum_{\substack{ i \in S\setminus T \\ t\in T}} (X_i - X_t)^2 ~+~ b~\sum_{\substack{ i \in S'\setminus T' \\ t\in T'}}(X_i - X_t)^2 \\
\end{align*}

From the condition in \prettyref{lem:suff-cond-lemma}, we get that the primal SDP is integral as long as $ bn \geq  8c/(1-\epsilon)$, which is true as long as $\frac{1-\epsilon}{\epsilon} > 8c$.  This is true for $\epsilon$ being less than a small enough constant. 
\end{proof}

We now consider the general case. We focus on just $S$ henceforth, as similar arguments will work for $S'$ too, and the feasible solution can be constructed independently for either part. Observe that in contrast to the complete graph above, certain terms are missing in the expansion of $X^T L_{\mid S}X$:   these terms are of the form 
\begin{equation*}
(X_i - X_t)^2  \qquad \qquad \forall i \in S\setminus T,\, t\in T: i \notin N(t)\mper
\end{equation*}

One way to recover these terms is to make use of the following observation:

\begin{fact}\label{fct:l-triang-ineq}
For any $x_1, x_2, \ldots x_{l+1} \in \R$, we have:

\[
\sum_{i=1}^{l} (x_i - x_{i+1})^2 \geq \frac{1}{l} (x_1 - x_{l+1})^2
\]
\end{fact}
\begin{proof}
Using Jensen's inequality, we have for any numbers $a_1, \ldots a_l \in \R$, $\frac{1}{n}\sum_i {a_i^2} \geq \paren{\frac{\sum_i a_i}{n}}^2$. Setting $a_i = x_i -x_{i+1}$ gives the required inequality.
\end{proof}
%
The above helps us restate our requirement as a flow routing problem.

\paragraph{Flow Routing:}
\prettyref{fct:l-triang-ineq} gives us a way to generate terms of the form $(X_i - X_t)^2$ using  the edges present in the graph $G[S]$. In particular, we can generate a missing term of the form $(X_i - X_t)^2$, as a sum along a path $P = (i_1 = i, i_2, i_3, \ldots, i_l=t)$ in $G$ of the terms $(X_{i_j} - X_{i_{j+1}})^2$, for every $j \in [l-1]$. Each of these terms occurs in the expansion of $X^T LX$. If we use an amount $a$  of the weight of each edge on $\mP$ in doing so, the final term has a coefficient of $\frac{a}{l}$, and this can be seen as $i$ attempting to sending a `flow' of magnitude $a$ to $t$ via $\mP$.

\smallskip
Generating all the missing terms can now be formulated as a flow-routing problem using paths of length at most $l$ (for some fixed $l$). The flows going from $i$ to $t$ generate the term $(X_i - X_t)^2$. \prettyref{lem:suff-cond-lemma} can therefore be restated as the problem of routing at least $c'l/n$ units of flow from every $i \in S \setminus T$ to $t \in T$.  The constraint on the (directed) flow edges out of $i$ is determined by the values $Y_{ij}$. The capacity of the edge $\set{i,j}$ in the direction $i\rightarrow j$ is $Y_{ij}$, and the outdegree constraint states that every vertex can push out at most one unit of flow in total. Furthermore, a flow of `$a$' units travelling along a path of distance $l$ to reach $t$ finally contributes only $a/l$, due to \prettyref{fct:l-triang-ineq}. We state this idea formally below.

\begin{lemma}[Flow routing problem] \label{lem:flow-routing-suffices}
Suppose we are given $G[S]$ and $G[S']$ with a feasible  assignment $Y$ for the edges. Consider a directed version of $G[S]$, where every edge $\set{i,j}\in E$ is replaced by the directed edges $(i,j)$ and $(j,i)$ with capacities $Y_{ij}$ and $Y_{ji}$ respectively. If for some $l \in \mathbb{N}$, and for every $i \in S\setminus T$ and $t\in T$, we can route a flow of $ c'l /n$  from $i \rightarrow t$ using paths of length at most $l$ in $G[S]$ (and similarly for $G[S']$), while obeying the (directed) capacity constraints on the edges, then we have:

\[X^T LX ~\geq~    \frac{c'}{n} \paren{ \sum_{i \in S \setminus T} \sum_{t \in T} 
		\paren{X_i - X_t}^2 + \sum_{i \in S' \setminus T'} \sum_{t \in T'} 
		\paren{X_i - X_t}^2} \qquad \forall X \in \R^n
\]
\end{lemma}

The proof follows easily given the preceding discussion, and hence we defer a formal proof to \prettyref{app:flow-routing-proof}.

\begin{remark}\label{rem:flow-lower-bound}
\textbf{A simple lower bound for the flow routing problem}:
The following argument shows that it is not enough to consider the flow routing problem alone, if we want to prove \prettyref{thm:sdpintegral} for all of the stated range of parameters. Suppose that the average distance between pairs of vertices in our graph $G[S]$ is $\mD$. For constant-degree edge-expanders, this is $\Omega(\log n)$. For a typical vertex $i$ which routes  $b$ units of flow to each $t \in T$, we would need $b  \geq \Omega(c'\mD/n)=  \Omega(\log n/n)$.
\smallskip
 However, since every node sends out $b \times \epsilon n$ units of flow ($b$ to each $t \in T$), the out-degree bound $\sum_j Y_{ij} \leq 1$ implies that $\epsilon n \times b \leq 1$, giving $b\leq 1/\epsilon n$. This contradicts the lower bound on $b$, unless $\epsilon \ll 1/\log n$.
\end{remark}

\smallskip

\begin{remark}\label{rem:expansion-lower-bound}
Consider the  case when $G[S]$ is a $d$-regular edge-expander with spectral gap $\lambda_1$. One could attempt to generate the missing terms using edge-expansion alone, as we have:
\[
X^T L _{\mid S}X = \sum_{ij \in E(G[S])} (X_i - X_j)^2 \geq \Omega \paren{\frac{\lambda_1 d}{n}} \sum_{ij \in S} (X_i - X_j)^2 \geq \Omega \paren{\frac{\lambda_1 d}{n}} \sum_{\substack{i \in S\setminus T \\ t \in T}} (X_i - X_t)^2
\] 

Although this gives us all the terms $(X_i - X_t)^2$, we have to weight all the edges uniformly (or close to it) by $b$, where $b\leq 1/d$.  The final coefficient of the term $(X_i - X_j)^2$ is therefore just $\lambda_1/ n$,  which is less than $c'/n$, since $c' \geq 4$. Furthermore, for $X$ defined as $X_i = 1$ for $i\in S \setminus T$, $0$ for $i \in T$, the second inequality is tight.  
\end{remark}

Thus, in order to prove our main result for the full range of parameters stated, we have to use a combination of the flow-routing technique and the edge-expansion properties of the graphs involved. With this in mind, we are now ready to complete the proof of the main theorem, which we restate here for clarity.

\begin{figure}
\centering
\label{fig:weights}
\includegraphics[scale=0.6]{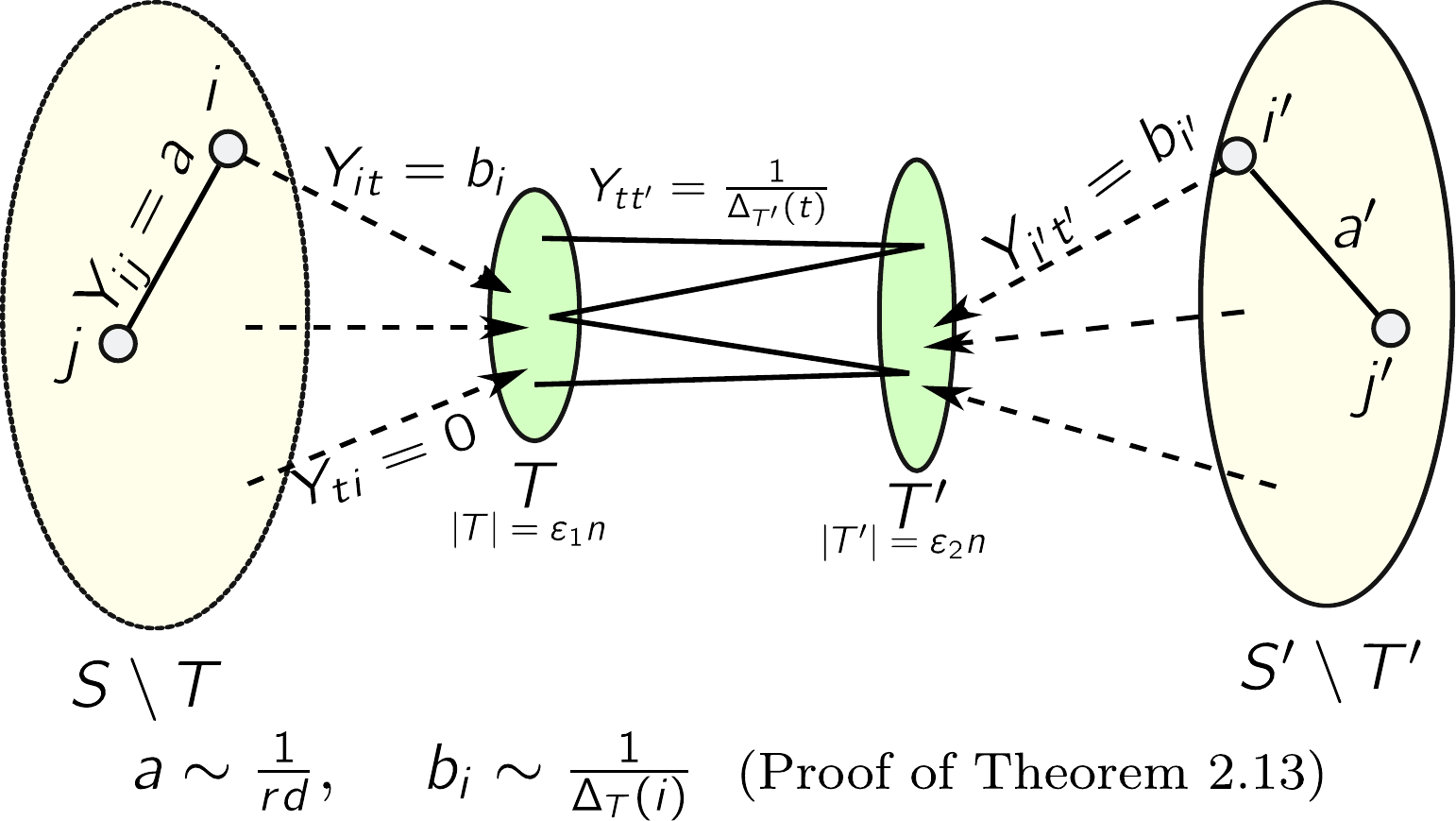}
	\caption{Outline of $Y_{ij}$'s used in the constructed dual solution.}
\end{figure}

\begin{theorem}[\prettyref{thm:sdpintegral} restated]\label{thm:main-thm-restated}
For the regime of parameters stated in \prettyref{thm:main}, 
$U' \defeq \inds \inds^T$ and for each $i$ $\eta_i' \defeq \max_{j \in N(i)} \paren{\inds(i) - \inds(j)}^2$ 
for the set $S$ defined in \vbmparams, 	is the unique optimal solution to \prettyref{sdp:primal}
with probabilty at least $1 - 1/{\sf poly}(n)$.
\end{theorem}

\begin{proof}
We will first use the expansion property of the edge-expander within $G[S\setminus T]$ (see \prettyref{prop:expansion-property}), and then route flows from $N_{S\setminus T}(T)$ to $T$. We will state our proof considering just $S$, a similar argument works for $S'$. For notational simplicity, let $\e = \e_1$, $p=p_1$, and $d$ be the minimum degree in the edge-expander graph within $G[S\setminus T]$.  As stated before, for edges ${i,j}$ not in the edge-expander, we set $Y_{ij} = Y_{ji}=0$. Hence,  without loss of generality, we may assume henceforth that $G[S \setminus T]$ is just the edge-expander graph, and ignore the other edges present
 
 \medskip
For every edge $\set{i,j}$ in $G[S\setminus T]$, we set $Y_{ij} = a$, and for every edge between $i \in N_{S\setminus T} (t)$ and $t \in T$, set $Y_{it}=b_i$; we will choose the values of $a$ and $b_i$'s appropriately later.  All other $Y_{ij}$'s are set to 0. 

\medskip
The following proposition follows from the spectral definition of edge-expander graphs. For completeness, we include a proof in \prettyref{sec:expansion-property}.

\begin{proposition}\label{prop:expansion-property}
Let $G=(V,E)$ be an edge-expander graph with Laplacian $L$, and spectral gap $\lambda$. Suppose that the degrees of the vertices in $G$ satisfy $\Delta(i) \in [d, rd]$, for some $r>1$, and $d \in \mathbb{N}$. Then for any $X \in \R^n$, we have:
\begin{equation}
\sum_{\set{i,j} \in E} (X_i - X_j)^2 ~\geq~ \frac{1}{r^2} \cdot \frac{\lambda d}{n} \sum_{i,j \in V \times V} (X_i - X_j)^2  
\end{equation}
\end{proposition}

\medskip
We first use the edge-expansion of $S\setminus T$ to effectively turn it into a complete graph via \prettyref{prop:expansion-property}.

\begin{equation} \label{eq:use-expansion}
 \sum_{i,j \in E(S\setminus T)} a  (X_i - X_j)^2 ~\geq~  \frac{ad\lambda_1}{r^2 n} \sum_{i,j \in S \setminus T} (X_i - X_j)^2 
 \end{equation}

\medskip
Consider a new graph $\hatG  \defeq (S, \widehat{E})$ on the vertex set $S$, with directed weights  on the edges (capacities) $\widehat{Y}$ as follows: $\hatG[S\setminus T]$ is a complete graph; we let $\hatY_{ij} = \hatY_{ji} = ad\lambda_1/r^2 n$ for every $i, j \in S\setminus T$. For every $i \in  S\setminus T $, $j \in T$, $\set{i,j} \in \widehat{E}$ iff $ \set{i,j} \in E$, and for such edges, let $\hatY_{ij}= Y_{ij}$. All other $\hatY_{ij}$'s are set to $0$.
\medskip

\begin{observation}
It suffices to route flows in $\hatG$ to satisfy \prettyref{lem:flow-routing-suffices}, in order to satisfy the conditions of \prettyref{lem:suff-cond-lemma} for $G$.
\end{observation}
\begin{proof}
Equation \prettyref{eq:use-expansion}, along with the definition of the capacities $\hatY$, immediately implies that
\[
X^T L(Y)_{\mid S} X ~\geq~ X^T L(\hatY)_{\mid S} X
\]
\end{proof}

The new capacities  for every directed edge in $\hatG[S\setminus T]$ are $ad\lambda_1/nr^2$. Note that this is a significant reduction from just $a$, which were the capacities in the original edges of the graph, but we gain in the presence of edges $\set{i,j}$, for every $j \in N_{S\setminus T}(T)$ and each $i \in S\setminus T$.  This allows us to route our flows in length-two steps: from $i$ to the neighbors of $t$, and from there on to $t$.

%
\medskip

We now define our flows on $\hatG$. For what follows, recall that by the definition of $\hatG$, $N_T(i)$ remains the same for a vertex $i\in S\setminus T$ in both $G$ and $\hatG$. Similarly, $N_{S\setminus T} (t)$ is unchanged for any vertex $t\in T$. 

\medskip

For any $t \in T$, consider an $i\in S\setminus (N(t) \cup T)$. Through every $j \in (N_{S\setminus T}(T) \setminus T)$, it sends $ad\lambda_1/(2r^2 n\Delta_T(j))$ to each $t \in N_T(j)$. These paths are of length at most $2$. We thus have that the flow routed from $i$ to $t$ is given by:

\begin{equation}\label{eq:flow-routed-general}
f(i\rightarrow t) ~=~ \frac{1}{2}\sum_{j \in N_{S\setminus T}(t)} \frac{ad\lambda_1}{2 r^2 n} \times \frac{1}{\Delta_T(j)} \qquad \qquad \text{(Flow routed)}
\end{equation}

The out-degree bound $\sum_{j \in N(i)} Y_{ij} \leq 1$ gives us (this should hold in the original graph $G[S \setminus T]$):

\begin{equation}\label{eq:outdeg-bound-general}
ard ~+~ b_i|\Delta_T(i)| \leq 1 \qquad \forall i \in S\setminus T \qquad \text{(Outdegree bound)}
\end{equation}

This is satisfied, if we set $a \seteq 1/(2 r d)$, and $b_i \seteq 1/(2\Delta_T(i))$. Note that if $\Delta_T(i)=0$, then it has no $b_i$-weighted outgoing edges.

\medskip

Finally, since every edge $j \rightarrow t$ in $S\setminus T \rightarrow T$ has to handle many units of flow, we need to have:

\begin{equation}\label{eq:b-bound-general}
b_j \geq \frac{ad\lambda_1}{2r^2 n\Delta_T(j)} \times (1-\epsilon)n \qquad \forall j\in N(T) \cap (S\setminus T) \mper
\end{equation}

Notice that from the setting for equation\eqref{eq:outdeg-bound-general}, we have that the above equation is, indeed, true.

\medskip
It remains to verify that the amount of flow that every $i \in S\setminus T$ sends to every $t\in T$ is large enough.  The amount of flow reaching is given by equation~\eqref{eq:flow-routed-general}:

\begin{equation}
f(i\rightarrow t) \geq \frac{ad\lambda_1}{4 r^2 n} \sum_{j \in N_{S\setminus T}(t)} \frac{1}{\Delta_T(j)} ~\geq~ \frac{\lambda_1}{8r^3n}\sum_{j \in N_{S\setminus T}(t)} \frac{1}{\Delta_T(j)}
\end{equation}

\medskip
It remains to lower-bound the sum $\sum_{j \in N(t)} \frac{1}{\Delta_T(j)}$, and we do this by using properties of the random graph between $S\setminus T$ and $T$. Suppose the following bounds on $\epsilon$ and $p$ hold, for some constant $C_1$: 
\begin{align*}
\epsilon \leq \frac{\lambda_1(1-\epsilon)}{384 \cdot C_1}, && p\geq 4\cdot \frac{64C_1 \log n}{\lambda_1 (1-\epsilon) n}. 
\end{align*}

Since the graph on $S\setminus T \times T$ is random, we have the following concentration bounds on the degrees of vertices in $T$ and $N_{S\setminus T}(T)$ respectively (recall that we set the weight of adversarial edges to zero, and hence don't consider them in this analysis):

\begin{Claim}\label{cl:deg-of-t}
For every $t \in T$, we have $\Delta_{S \setminus T}(t) \in [p(1-\epsilon) n/4, 2 p (1-\epsilon)n]$  with probability at least $1-\frac{1}{n^3}$. 
\end{Claim}
\begin{proof}
The expected degree $\Delta_{S\setminus T}(t)$ of any vertex $t \in T$ is $p(1-\eps)n/2 = \Omega(\log n)$. The result follows from a direct application of the Chernoff bound on the probability that any of the degrees deviate from their expectation.
\end{proof}

\begin{Claim}\label{cl:deg-of-n-t}

Let $M = \frac{\lambda_1 p (1-\epsilon) n}{64 C_1}$. Then:
\[
\Pr{\forall i \in S\setminus T ~:~ \Delta_T(i) \leq M} ~\geq~ 1- n\cdot 2^{-M} ~\geq~ 1- \frac{1}{n^3} 
\]
\end{Claim}

\begin{proof}
We will need the following version of the Chernoff bound:

\begin{theorem}[{{\cite[Theorem 4.4 Chapter 4]{MU05}}}]\label{thm:large-dev-chernoff}
Let $X_1, \ldots X_n$ be independent $0-1$ random variables, and let $X=\sum_i X_i$, and $\mu= \E[X]$. Then for $R\geq 6\mu$,
\[
\Pr{X \geq R} \leq 2^{-R}
\]
\end{theorem}

By our bounds on $\epsilon$ and $p$, we get the following lower bounds on $M$ respectively:

\[
M ~\geq~ 6 p \epsilon n, \qquad \qquad M ~\geq~ 4 \log n
\] 

Let $X^i_t$ be the indicator random variable, which is $1$, if an edge exists between $i\in S\setminus T$ and $t\in T$, and $0$ otherwise. We have $\Delta_T(i)= \sum_{t \in N_T(i)} X^i_t$. A direct application of the Chernoff bound in \prettyref{thm:large-dev-chernoff} to the random variable $\Delta_T(i)$, followed by a union bound over $i$, yields the statement of the claim.
\end{proof}

With this uniform bound on $\Delta_T(i)$ that holds whp, we have that the flow from every $i \in S \setminus T$ to $t \in T$ is (whp) at least:
\begin{align*}
f(i\rightarrow t) &~\geq~ \frac{\lambda_1}{8r^3 n} \sum_{j \in N_{S\setminus T}(t)} \frac{1}{\Delta_T(j)} \\
& ~\geq~  \frac{\lambda_1}{8 r^3 n} \times \frac{1}{M} \times \frac{p(1-\epsilon)n}{4} \qquad &\text{\ldots  using  Claim~\ref{cl:deg-of-t}} \\
& ~\geq~  2 C_1 \times \frac{1}{r^3 n} \qquad &\text{\ldots substituting for $M$}
\end{align*}

Since we needed to get $2c'/n$ units of flow from $i$ to $t$, choosing $C_1 \geq r^3 c'$ is sufficient.
For $\lambda_1$ being a constant, the bounds satisfy the conditions in the theorem statement. This completes the proof of \prettyref{thm:main-thm-restated}.
\end{proof}

\begin{remark}
It is instructive to note that the above assignment for $Y$ (and in fact, any assignment satisfying the conditions in \prettyref{lem:suff-cond-lemma},) sets $ Y_{tj} =0$, for every $t \in T$, and $j \in S$, which is consistent with complementary slackness conditions on the optimal dual solutions, if the primal solution is  $\one_S \one_S^T$.  Such an assignment also effectively decouples the setting of edge weights $Y$  in $S$, $S'$ and the bipartite subgraph on $T \times T'$.
\end{remark}

\begin{remark}\label{rem:main-holds-for-general}
An interesting parameter in the above proof is the harmonic sum of certain degrees associated with every boundary vertex. For a boundary vertex $t \in T$ (and similarly for $t' \in T'$), define the quantities
\[
H_{\text{ext}}(t) \eqdef \sum_{t'  \in N_{T'} (t)} \frac{1}{\Delta_T (t')} \mcom\qquad 
H_{\text{int}}(t) \eqdef \sum_{j \in N_{S\setminus T}(t)} \frac{1}{\Delta_T(j)} \mcom
\]
as the \emph{external} and \emph{internal} harmonic sums respectively.  Our proof in fact shows that the following more general theorem holds:

\begin{theorem}[\prettyref{thm:sdpintegral}, generalized]\label{thm:sdpintegral-general}
For $G$ in \vbmparams,  \prettyref{sdp:primal} is integral if :
$$ H_{\mathrm{int}}(t) ~\geq~ \alpha(c, r, \lambda_1, \lambda_2) \cdot   H_{\mathrm{ext}}(t) \qquad  \forall t\in T \cup T'  \mcom$$

Above, $\alpha$ is a constant depending on the parameters $c,r, \lambda_1, \lambda_2$ of the model.
\end{theorem} 

\noindent In our proof, the randomness in \vbm~was used to show that the above bound holds with high probability.

\end{remark}

\begin{remark}
Consider a set of vectors $u_1, \ldots u_n \in \R^n$, such that $\norm{u_i}^2 =1$, and $u_i^Tu_j = -\frac{1}{n-1}$.  Such a set of vectors are known to exist~\cite{kms98}. The pair $(U,\eta)$, with $U_{ij} = u_i^T u_j$ and $\eta_i = 2+2/(n-1)$ constitute a feasible primal solution of objective value $2n(1+1/(n-1))$, for \emph{any} graph $G$.  This shows that \prettyref{sdp:primal} is not integral for the \vbm~ model, if $\e_1 + \e_2 \geq \frac{1}{2} (1+o(1))$.  
\end{remark}

\newcommand{\hatvbm}{\widetilde{\vbm}}
\newcommand{\hatvbmparams}{$\widetilde{\vbm}(n, \e, p)$}
\newcommand{\planted}{\textsf{Planted}}

\section{Constant-factor approximations in related models} \label{sec:const-factor-approx}

We now  give constant-factor bi-criteria approximations for finding a balanced cut with vertex expansion close to that of the planted cut.   These correspond to allowing a broader range of  the parameters $p_i$ and $\lambda_i$ in the definition of $\vbm$. While one allows the edges between $S \setminus T$ and $T$ to be completely adversarial, the other allows the graph on $S\setminus T$ to be arbitrary.

\medskip 
For our algorithms, we will refer to the SDP solution in terms of the vectors $\set{u_i}_{i \in V}$ obtained by factorizing the SDP solution $U$. These satisfy $u_i^T u_j = U_{ij}$. We will strengthen \prettyref{sdp:primal} by adding in the $\ell_2^2$ triangle inequality constraints, most conveniently stated in terms of the vectors $u_i$:

\[
\forall i,j,k~:~ \norm{u_i - u_j}^2 + \norm{u_j - u_k}^2 \geq \norm{u_i - u_k}^2
\]

It is not difficult to see that these constraints are satisfied by any integral solution. 

\smallskip

In both our models,  we show that the set of SDP solution vectors has a \emph{dense cluster} of $\Omega(n)$ vertices, lying within a ball of small $\ell_2^2$ diameter. Using this with the balance constraint and a \emph{line-embedding} of the vertices gives us our results. 
We give below a lemma formalizing how existence of a dense cluster can be used to recover an almost balanced cut with small vertex expansion. We leave the proof to \prettyref{app:cluster-suffices-proof}, since it uses arguments that are standard in current literature.

%

\begin{lemma}\label{lem:cluster-suffices}
Given the optimal SDP solution vectors $\{u_i\}_{i \in V}$ with objective value $\delta n$ to \prettyref{sdp:primal} strengthened by the $\ell_2^2$ triangle inequalities for an instance $G$, if there exists a set $L$ that satisfies: 
\begin{enumerate}
\item[(a)] $\abs{L} \geq \alpha n$
\item[(b)] $\sum_{i, j \in L} \norm{u_i - u_j}^2 \leq \frac{1}{40} \cdot \abs{L}^2 $
\end{enumerate}
for some constant $\alpha$, then we can recover, using Algorithm 1, in polynomial time, a balanced partition $(W, W')$ with $\abs{W}, \abs{W'} \geq  \Omega(\alpha n)$ and $\phiv(W)=O(\delta/\alpha)$.
\end{lemma}

\begin{algorithm} \label{alg:vexp-alg-2}
\caption{Algorithm for rounding clustered SDP solutions}
\begin{algorithmic}[1]
\INPUT $G=(V,E)$ and an optimal SDP solution  $\inbrace{u_i}_{i\in V}$ on $G$, a parameter $\alpha$.
\OUTPUT $W^* \subseteq V$ with $\abs{W^*} \in [3\alpha n/4, 14n/15]$.
\For{$i \in V$}
    \State Let $L'_i \eqdef \set{j~:~ \norm{u_i - u_j}^2 \leq 1/10}$. Let $j^*(i) = \arg \min_{j \in V :  \norm{u_i - u_j}^2> 1/8} \paren{\norm{u_i - u_j}^2 - 1/8}$ .
    \State If $\abs{L'_i} \leq 3\alpha n /4$, \textbf{continue}.
    \State Sort the points $u_j$ in increasing order of squared-distances: $d(j,L'_i) \eqdef \min_{k\in L'_i} \norm{u_j - u_k}^2$. Denote the corresponding ordered points as $j_1, \ldots j_n$.
     \State Let $S_l \eqdef \{ j_1, \ldots j_l\}$. Set $W^*_i \; \gets \; \arg\min_{l < j^*(i)} \,\,\,  \phiv(S_l)$.
\EndFor
\State Let  $ i^* = \arg\min_i \phiv(W^*_i)$. Output $W^*_{i^*}$.
\end{algorithmic} 
\end{algorithm}

We will also be using the following easy lemma:

\begin{lemma}\label{lem:min-max-degree}
 For a graph $G=(V,E)$ with minimum and maximum degrees $\Delta_{\min}$ and $\Delta_{\max}$ respectively,  let $f(i,j): V\times V \rightarrow \R^{\geq 0}$ be some non-negative symmetric function on pairs of vertices. Then,
\[
\E_{i \in V} \insquare{\max_{j \in N(i)} f(i,j)} \geq \paren{\frac{\Delta_{\min} }{\Delta_{\max}}} \E_{\set{i,j}\in E(G) } f(i,j)
\]

If $G$ is bipartite with bipartition $L, R$, then we further have:  
\[\E_{i \in L} \insquare{\max_{j \in N(i)} f(i,j)} \geq \paren{\frac{\Delta_{\min}(L) }{\Delta_{\max} (L)}} \E_{\set{i,j}\in E(G) } f(i,j)
\]
\end{lemma}
\begin{proof}
\begin{align*}
\E_{i\in V}\insquare{\max_{j \in N(i)} f(i,j)} &= \frac{1}{|V|}\sum_{i \in V} \max_{j \in N(i)} f(i,j)
~\geq~ \frac{1}{|V|}\sum_{i \in V} \frac{1}{\Delta(i)}\sum_{j \in N(i)} f(i,j) \\
&\geq~ \frac{2}{|V| \Delta_{\max}} \sum_{\set{i,j} \in E(G)} f(i,j) \\
& \geq \paren{\frac{2|E|}{|V| \Delta_{\max}}} \E_{\set{i,j} \in E(G)} \insquare{f(i,j)} ~\geq~  \paren{\frac{\Delta_{\min} }{\Delta_{\max}}} \E_{\set{i,j}\in E(G) } f(i,j)
\end{align*}

For the bipartite case, the proof is almost the same, with $V$ replaced by $L$ in each of the steps above. The only difference is that there is no loss of the factor of $2$ in the second inequality, as every edge is counted exactly once (with $\Delta_{\min}(L)$ and $\Delta_{\max}(L)$ being the minimum and maximum degrees in $L$ alone).
\end{proof}

\subsection{When $S\setminus T$ is an arbitrary graph} \label{sec:vbmlambda}

In the $\vbm$ model and the corresponding proof of the integrality of the SDP (\prettyref{thm:sdpintegral}), we used the fact that the subset $S\setminus T$ and $S' \setminus T'$ both contain an almost-regular spanning expander as a subgraph. If we relax this constraint (i.e, consider the model \vbmparamslambda), we can still find a bi-criteria approximation. (Note that the parameter $r$ is of no consequence when $\lambda_i=0$).

%
%
%
%
%
%

\begin{theorem}[\prettyref{thm:vbmlambda} restated]\label{thm:vbmlambda-main}
There exist universal constants 
$c_1 \in \R^+$, $c_2 \in (0, 1/2), c_6 \in (0,1/2)$ satisfying the following:	
there exists a polynomial time algorithm which takes a graph generated from \vbmparamslambda,
where $\e_1,\e_2 \in [1/n, c_2]$ and $p_1 (\e_1 + \e_2) n \geq c_1 \log n$,
and outputs  with probability at least $1 - 1/{\sf poly}(n)$, 	
a set $A \subset V$ satisfying $\Abs{A} \in [ c_6 n, (1 - c_6) n]$
and $\phiv(A) \leq \bigo{\e_1 + \e_2}$.
\end{theorem}


\begin{proof}
The following proposition will help us show that for the given range of $p$, the vectors of $S\setminus T$ are clustered. We defer the proof to \prettyref{app:bipartite-expansion-section}.

\begin{proposition}
\label{prop:br-1}
Let $L,R$ be two (disjoint) sets of vertices of sizes $(1 - \gamma) n$ and $\gamma n$ respectively, where $\gamma \leq 1/2$.
Let $p \in (0,1)$ be any number satisfying $p \gamma n \geq 300 \log n$.	
Let $\tilde{G}$ be a random bipartite graph obtained by adding edges between each pair in $L \times R$
independently with probability $p$. 
Let $u_1, \ldots, u_n$ be a set of vectors. Then, there exists an absolute constant $C$ such that, with high probability, we have:
\begin{enumerate}
 \item[a)] \[ \E_{i,j \in L} \norm{u_i - u_j}^2 \leq C \E_{\set{ij} \in E(\tilde{G})} \norm{u_i - u_j}^2  \mper  \]
\item[b)]The  minimum and maximum degrees $\Delta_{\min}(L),~\Delta_{\max}(L)$ in $L$ satisfy $\Delta_{\max}(L) / \Delta_{\min}(L)  \leq 3$.
\end{enumerate}

\end{proposition}

\medskip

Let $H=(S\setminus T, T, E(H))$ be the random bipartite graph between $S\setminus T$ and $T$ (ignoring any of the monotone adversarial edges added in what follows, as the added edges would only help in the inequalities). We apply \prettyref{prop:br-1} to $H$, and condition on the events mentioned happening.
\medskip

Suppose that the optimal SDP solution $(U, \eta)$ has value $\delta n$; we know that $\delta \leq (\eps_1+\eps_2)/4$ . Let vectors $\{u_i\}_{i \in V}$ be obtained by factorizing the SDP solution $U$. We know that:

$$\sum_{i \in V} \eta_i \leq \delta n ~\implies~ \sum_{ i \in S \setminus T} \eta_i  \leq \delta n$$

Thus, we have that $\E_{ i \in S \setminus T} [\eta_i] \leq \delta /(1/2-\e_1) \leq 4\delta $. Now, 

\begin{align*}
\E_{i,j \in S\setminus T} \norm{u_i - u_j}^2 &\leq~ C \E_{\set{ij} \in E(H)} \norm{u_i - u_j}^2  \\
&\leq~ C \cdot \paren{\frac{\Delta_{\max} (S \setminus T)}{\Delta_{\min}(S \setminus T)}} \cdot \E_{i \in S\setminus T}\insquare{ \max_{j \in N_T(i)}\norm{u_i - u_j}^2} && \ldots \text{Using \prettyref{lem:min-max-degree} on $H$ with $L = S \setminus T$}\\
& \leq~ 3C \E_{i \in S\setminus T}\insquare{ \max_{j \in N_T(i)}\norm{u_i - u_j}^2} && \ldots \text{Using \prettyref{prop:expansion-property} (b)}  \\
& \leq~ 3C \E_{i \in S\setminus T}\insquare{ \max_{j \in N(i) }\norm{u_i - u_j}^2} \\
&\leq 12 C \delta \leq \frac{1}{40} && \ldots \text{If $\eps_1 + \eps_2 \leq 1/(120 C)$}
\end{align*}

The last step can be ensured for $c_2 \leq 1/120 C$. We now invoke \prettyref{lem:cluster-suffices} on the set $S\setminus T$ with $\alpha = (1/2 - \eps_1)$ to complete the proof.

%
%
%
%

\end{proof}

\subsection{VBM without randomness} \label{sec:vbmp}
When the graph between $S\setminus T$ and $T$ (and between $S'\setminus T'$ and $T'$) is arbitrary, it corresponds to setting $p_1$ and $p_2$ to zero in \vbmparams. The graph is no longer random; in this case we show that, we can still get a good partition (for every instance in this class). In fact, we will prove the following slightly more general theorem, where all we require is that $S$ contains a close-to-regular linear-sized expander.

\begin{theorem}\label{thm:vbmp-main}
Let $G=(V,E)$ be a graph with a  planted bisection $(S, S')$ having $\eps n$ boundary vertices. Suppose $S$ contains a spectral expander $H=(M, E(H))$ on $M \subseteq S$ as a subgraph, with spectral gap $\lambda$,   $\abs{M} \geq \beta n$, and $\Delta_{\max}(H) /\Delta_{\min}(H) \leq r$. If $\lambda \geq 320 \eps r^3/\beta$,  we can find  a partition $(A, V\setminus A)$ with $\abs{A}, \abs{V\setminus A} = \Omega(n)$ and  $\phiv(A)=\bigo{\eps}$.
\end{theorem}

\begin{proof}
Let $\{u_i\}_{i\in V}$ be the SDP solution vectors with objective value $\delta n$, we know that $\delta \leq 4\eps$. Since $\E_{i\in V}[ \max_{j \in N(i)}\norm{u_i - u_j}^2] = \delta$, we should have that  $\E_{i \in M}[ \max_{j \in N(i)}\norm{u_i - u_j}^2] \leq \delta/\beta$. Consider the subgraph $H$ on $M$; we have :

\begin{equation}
\E_{i \in M}\insquare{ \max_{j \in N(i)\cap M}\norm{u_i - u_j}^2} \leq \E_{i \in S\setminus T} \insquare{ \max_{j \in N(i)}\norm{u_i - u_j}^2 } \leq \delta/\beta
\end{equation}  

\smallskip
We can now exploit the spectral expansion property of $H$.

\begin{align*}
\E_{i \in M, j \in M}\insquare{\norm{u_i - u_j}^2} & \leq \frac{r^2}{\lambda}\E_{\set{ij} \in E(H)} \insquare{\norm{u_i - u_j}^2} \qquad \qquad \ldots \text{ Using \prettyref{prop:expansion-property} on $H$ }\\
&\leq \paren{\frac{r^3}{\lambda}}~\E_{i \in M}\insquare{\max_{j \in N(i) \cap M}\norm{u_i - u_j}^2} \qquad \ldots \text{ Using \prettyref{lem:min-max-degree} with $H$}\\
&\leq \frac{r^3\delta}{\beta\lambda} \leq \frac{1}{40}
\end{align*}

Invoking \prettyref{lem:cluster-suffices} on the set $M$ would now give us the required result.
\end{proof}

The proof of \prettyref{thm:vbmp} follows almost immediately:

\begin{proof}[Proof Of \prettyref{thm:vbmp}]
Invoking \prettyref{thm:vbmp-main} with $\beta = (1/2-\eps_1)$,  and $M = S\setminus T$, since $\lambda_1$ satisfies the required conditions, we can find in polynomial time a set $A$ with $\phiv(A) \leq  \bigO(\eps_1+\eps_2)$, and $A, V\setminus A$ being of size $\Omega(n)$.
\end{proof}

\begin{remark}
The \emph{planted model} considered by \cite{lr14} (which we describe in \prettyref{app:lr14}) also falls in the class described above, and we therefore obtain a constant-factor bi-criteria approximation for instances in it. The \cite{lr14} planted model is a direct analogue of the planted model in \cite{mmv12}. 

\end{remark}

\paragraph*{Acknowledgements.}
We thank Amit Deshpande for many helpful discussions. Rakesh Venkat was visting Microsoft Research, Bangalore when this research was initiated. Anand Louis is grateful to MSR, Bangalore for supporting this collaboration.

{\small
\bibliographystyle{alpha}
\bibliography{vertexexp-bib}

\begin{thebibliography}{ABKK17}

\bibitem[Abb17]{a17}
Emmanuel Abbe.
\newblock Community detection and stochastic block models: recent developments.
\newblock {\em arXiv preprint arXiv:1703.10146}, 2017.

\bibitem[ABBS14]{abbs2014}
Emmanuel Abbe, Afonso~S Bandeira, Annina Bracher, and Amit Singer.
\newblock Decoding binary node labels from censored edge measurements: Phase
  transition and efficient recovery.
\newblock {\em IEEE Transactions on Network Science and Engineering},
  1(1):10--22, 2014.

\bibitem[ABH16]{abh16}
Emmanuel Abbe, Afonso~S Bandeira, and Georgina Hall.
\newblock Exact recovery in the stochastic block model.
\newblock {\em IEEE Transactions on Information Theory}, 62(1):471--487, 2016.

\bibitem[ABKK15]{abkk15}
Naman Agarwal, Afonso~S Bandeira, Konstantinos Koiliaris, and Alexandra Kolla.
\newblock Multisection in the stochastic block model using semidefinite
  programming, 2015.
\newblock arXiv preprint arXiv:1507.02323. {\em To Appear in Compressed Sensing
  and Its Applications: Second International MATHEON Conference}.

\bibitem[ABKK17]{abkk2017}
Naman Agarwal, Afonso~S Bandeira, Konstantinos Koiliaris, and Alexandra Kolla.
\newblock Multisection in the stochastic block model using semidefinite
  programming.
\newblock In {\em Compressed Sensing and its Applications}, pages 125--162.
  Springer, 2017.

\bibitem[Alo86]{a85}
Noga Alon.
\newblock Eigenvalues and expanders.
\newblock {\em Combinatorica}, 6(2):83--96, 1986.

\bibitem[AM85]{am85}
Noga Alon and Vitali~D Milman.
\newblock $\lambda_1$, isoperimetric inequalities for graphs, and
  superconcentrators.
\newblock {\em Journal of Combinatorial Theory, Series B}, 38(1):73--88, 1985.

\bibitem[ARV09]{arv09}
Sanjeev Arora, Satish Rao, and Umesh~V. Vazirani.
\newblock Expander flows, geometric embeddings and graph partitioning.
\newblock {\em Journal of the ACM}, 56(2), 2009.
\newblock (Preliminary version in {\em 36th STOC}, 2004).

\bibitem[AS15a]{as15a}
Emmanuel Abbe and Colin Sandon.
\newblock Community detection in general stochastic block models: Fundamental
  limits and efficient algorithms for recovery.
\newblock In {\em IEEE 56th Annual Symp. on Foundations of Computer Science
  (FOCS), 2015}, pages 670--688. IEEE, 2015.

\bibitem[AS15b]{as15b}
Emmanuel Abbe and Colin Sandon.
\newblock Recovering communities in the general stochastic block model without
  knowing the parameters.
\newblock In {\em Advances in neural information processing systems}, pages
  676--684, 2015.

\bibitem[AS17]{as17}
Emmanuel Abbe and Colin Sandon.
\newblock Detection in the stochastic block model with multiple clusters: proof
  of the achievability conjectures, acyclic bp, and the information-computation
  gap, 2017.
\newblock arXiv: 1512.09080 To Appear in Communications on Pure and Applied
  Mathematics (2017).

\bibitem[BHT00]{bht00}
Sergey Bobkov, Christian Houdr{\'e}, and Prasad Tetali.
\newblock {$\lambda_\infty$, Vertex Isoperimetry and Concentration}.
\newblock {\em Combinatorica}, 20(2):153--172, 2000.

\bibitem[Bop87]{b87}
Ravi~B. Boppana.
\newblock Eigenvalues and graph bisection: An average-case analysis.
\newblock In {\em Proceedings of the 28th Annual Symposium on Foundations of
  Computer Science}, SFCS '87, pages 280--285, Washington, DC, USA, 1987. IEEE
  Computer Society.

\bibitem[BP99]{bm99}
Roberto Battiti and Marco Protasi.
\newblock Approximate algorithms and heuristics for max-sat.
\newblock In {\em Handbook of Combinatorial Optimization: Volume1--3}, pages
  77--148, Boston, MA, 1999. Springer US.

\bibitem[BV04]{boydV2004}
Stephen Boyd and Lieven Vandenberghe.
\newblock {\em Convex Optimization}.
\newblock Cambridge University Press, New York, NY, USA, 2004.

\bibitem[CLTZ18]{cltz18}
T.{-}H.~Hubert Chan, Anand Louis, Zhihao~Gavin Tang, and Chenzi Zhang.
\newblock {Spectral Properties of Hypergraph Laplacian and Approximation
  Algorithms}.
\newblock {\em J. {ACM}}, 65(3):15:1--15:48, 2018.

\bibitem[CO07]{coja2007}
Amin Coja-Oghlan.
\newblock Colouring semirandom graphs.
\newblock {\em Combinatorics, Probability and Computing}, 16(4):515–552,
  2007.

\bibitem[FHL08]{fhl08}
Uriel Feige, MohammadTaghi Hajiaghayi, and James~R. Lee.
\newblock Improved approximation algorithms for minimum weight vertex
  separators.
\newblock {\em SIAM Journal on Computing}, 38(2):629--657, 2008.

\bibitem[FK01]{fk01}
Uriel Feige and Joe Kilian.
\newblock Heuristics for semirandom graph problems.
\newblock {\em Journal of Computer and System Sciences}, 63(4):639--671, 2001.

\bibitem[GV16]{gv16}
Olivier Gu{\'e}don and Roman Vershynin.
\newblock Community detection in sparse networks via grothendieck’s
  inequality.
\newblock {\em Probability Theory and Related Fields}, 165(3-4):1025--1049,
  2016.

\bibitem[HLL83]{hll83}
Paul~W Holland, Kathryn~Blackmond Laskey, and Samuel Leinhardt.
\newblock Stochastic blockmodels: First steps.
\newblock {\em Social networks}, 5(2):109--137, 1983.

\bibitem[HWX16]{hwx2016}
Bruce Hajek, Yihong Wu, and Jiaming Xu.
\newblock Achieving exact cluster recovery threshold via semidefinite
  programming: Extensions.
\newblock {\em IEEE Transactions on Information Theory}, 62(10):5918--5937,
  2016.

\bibitem[JS98]{js98}
Mark Jerrum and Gregory~B Sorkin.
\newblock The metropolis algorithm for graph bisection.
\newblock {\em Discrete Applied Mathematics}, 82(1):155--175, 1998.

\bibitem[KBG17]{kbg17}
Chiheon Kim, Afonso~S. Bandeira, and Michel~X. Goemans.
\newblock Community detection in hypergraphs, spiked tensor models, and
  sum-of-squares.
\newblock {\em arXiv preprint: arXiv:1705.02973 [cs.DS]}, 2017.

\bibitem[KK95]{kk95}
George Karypis and Vipin Kumar.
\newblock Analysis of multilevel graph partitioning.
\newblock In {\em Proceedings of the 1995 ACM/IEEE Conference on
  Supercomputing}, Supercomputing '95, New York, NY, USA, 1995. ACM.

\bibitem[KK98]{kk98}
George Karypis and Vipin Kumar.
\newblock A fast and high quality multilevel scheme for partitioning irregular
  graphs.
\newblock {\em SIAM J. Sci. Comput.}, 20(1):359--392, December 1998.

\bibitem[KM72]{km72}
Victor Klee and George~J Minty.
\newblock How good is the simplex algorithm.
\newblock In {\em Shisha, Oved. Inequalities III (Proc. of 3rd Symp. on
  Inequalities, UCLA)}, pages 159--175. Academic Press, California, 1972.

\bibitem[KMS98]{kms98}
David Karger, Rajeev Motwani, and Madhu Sudan.
\newblock Approximate graph coloring by semidefinite programming.
\newblock {\em J. ACM}, 45(2):246--265, March 1998.

\bibitem[LM14]{lm14b}
Anand Louis and Yury Makarychev.
\newblock {Approximation Algorithms for Hypergraph Small Set Expansion and
  Small Set Vertex Expansion}.
\newblock In {\em Approximation, Randomization, and Combinatorial Optimization.
  Algorithms and Techniques (APPROX/RANDOM 2014)}, volume~28 of {\em Leibniz
  International Proceedings in Informatics (LIPIcs)}, pages 339--355, 2014.

\bibitem[LR99]{lr99}
Tom Leighton and Satish Rao.
\newblock Multicommodity max-flow min-cut theorems and their use in designing
  approximation algorithms.
\newblock {\em J. ACM}, 46(6):787--832, November 1999.

\bibitem[LR14]{lr14}
Anand Louis and Prasad Raghavendra, 2014.
\newblock Personal Communication.

\bibitem[LRV13]{lrv13}
Anand Louis, Prasad Raghavendra, and Santosh Vempala.
\newblock The complexity of approximating vertex expansion.
\newblock In {\em Proc. of the 54th Annual Symp. on Foundations of Computer
  Science}, FOCS '13, pages 360--369, Washington, DC, USA, 2013. IEEE Computer
  Society.

\bibitem[Mas14]{m14}
Laurent Massouli{\'e}.
\newblock Community detection thresholds and the weak ramanujan property.
\newblock In {\em Proc. of the 46th Annual ACM Symp. on Theory of Computing},
  STOC '14, pages 694--703, New York, NY, USA, 2014. ACM.

\bibitem[McS01]{m01}
Frank~D. McSherry.
\newblock Spectral partitioning of random graphs.
\newblock In {\em Proc. of the 42nd IEEE Symp. on Foundations of Computer
  Science (FOCS)}, pages 529--537, Washington, DC, USA, 2001. IEEE Computer
  Society.

\bibitem[MMV12]{mmv12}
Konstantin Makarychev, Yury Makarychev, and Aravindan Vijayaraghavan.
\newblock Approximation algorithms for semi-random partitioning problems.
\newblock In {\em Proc. of the 44th Annual ACM Symp. on Theory of Computing},
  STOC '12, pages 367--384. ACM, 2012.

\bibitem[MMV14]{mmv14}
Konstantin Makarychev, Yury Makarychev, and Aravindan Vijayaraghavan.
\newblock Constant factor approximation for balanced cut in the pie model.
\newblock In {\em Proc. of the 46th Annual ACM Symp. on Theory of Computing},
  STOC '14, pages 41--49, New York, NY, USA, 2014. ACM.

\bibitem[MMV16]{mmv16}
Konstantin Makarychev, Yury Makarychev, and Aravindan Vijayaraghavan.
\newblock Learning communities in the presence of errors.
\newblock In {\em 29th Annual Conference on Learning Theory}, volume~49 of {\em
  Proceedings of Machine Learning Research}, pages 1258--1291, Columbia
  University, New York, New York, USA, 23--26 Jun 2016. PMLR.

\bibitem[MNS14]{mns14a}
Elchanan Mossel, Joe Neeman, and Allan Sly.
\newblock Belief propagation, robust reconstruction and optimal recovery of
  block models.
\newblock In {\em Conference on Learning Theory}, pages 356--370, 2014.

\bibitem[MNS15]{mns15a}
Elchanan Mossel, Joe Neeman, and Allan Sly.
\newblock Consistency thresholds for the planted bisection model.
\newblock In {\em Proc. of the 47th Annual ACM Symp. on Theory of Computing},
  STOC '15, pages 69--75, New York, NY, USA, 2015. ACM.

\bibitem[MNS17]{mns15b}
Elchanan Mossel, Joe Neeman, and Allan Sly.
\newblock A proof of the block model threshold conjecture.
\newblock {\em Combinatorica}, 2017.

\bibitem[MPW16]{mpw16}
Ankur Moitra, William Perry, and Alexander~S Wein.
\newblock How robust are reconstruction thresholds for community detection?
\newblock In {\em Proceedings of the forty-eighth annual ACM symposium on
  Theory of Computing}, pages 828--841. ACM, 2016.

\bibitem[MU05]{MU05}
Michael Mitzenmacher and Eli Upfal.
\newblock {\em Probability and Computing: Randomized Algorithms and
  Probabilistic Analysis}.
\newblock Cambridge University Press, New York, NY, USA, 2005.

\bibitem[RS10]{rs10}
Prasad Raghavendra and David Steurer.
\newblock Graph expansion and the unique games conjecture.
\newblock In {\em Proceedings of the Forty-second ACM Symposium on Theory of
  Computing}, STOC '10, pages 755--764, New York, NY, USA, 2010. ACM.

\end{thebibliography}
}

\IfFileExists{./appendix.tex}{\appendix

\newcommand{\hsym}{\phi^{\sf V}}
\section{Omitted Proofs}

\subsection{SDP Duality}
\label{app:dual}
In this section we show how to derive the dual of \prettyref{sdp:primal}. For more details on this, we refer the reader to \cite[Chapter 5]{boydV2004}. We first start by defining the Lagrangian dual $D(\eta,U,Y,B,\alpha)$ of \prettyref{sdp:primal},
using $\set{Y_{ij}}_{i,j}$, $\set{B_{ii}}_i$ and $\alpha$ as the dual variables.

\begin{align*} 
D(\eta,U,Y,B,\alpha) & \defeq \sum_{i \in V} \eta_i + \sum_{i \in V} \sum_{j \in N(i)} 
	\paren{U_{ii} + U_{jj} - 2 U_{ij} - \eta_i} Y_{ij}
	+ \sum_{i \in V} \paren{1 - U_{ii}} B_{ii} + 
	\alpha \sum_{i \in V} \sum_{j \in V} U_{ij} \\
	& = \sum_{i \in V} B_{ii} + \sum_{i \in V} \eta_i \paren{1 - \sum_{j \in N(i)} Y_{ij}}
	+ \sum_{i \in V} U_{ii} \sum_{j \in N(i)} \paren{Y_{ij} + Y_{ji} - B_{ii}} \\ 
	& \qquad - 2 \sum_{\set{i,j} \in E} U_{ij} \paren{Y_{ij} + Y_{ji}}
	+ \alpha \paren{\one \one^T} \cdot U \\
	& = \sum_{i \in V} B_{ii} + \sum_{i \in V} \eta_i \paren{1 - \sum_{j \in N(i)} Y_{ij}}
	+ \paren{L(Y) + \alpha \one \one^T - B} \cdot U \\ 
\end{align*}
Now, 
\[ \inf_{\substack{\eta,U \\U \succeq 0}} D(\eta,U,Y,B,\alpha) = \begin{cases}
	\sum_{i \in V} B_{ii} & \sum_{j \in N(i)} Y_{ij} = 1 \ \forall i \in V, \textrm{ and } 	
					L(Y) + \alpha \one \one^T - B \succeq 0, B_{ij} = 0\ \forall i \neq j \\
		- \infty & \textrm{otherwise} 
	\end{cases} \mper 
\]
Therefore, we get that the dual of \prettyref{sdp:primal} is
\begin{flalign*}
&\max \sum_{i \in V} B_{ii} \qquad \subjectto \\ 
&\sum_{j \in N(i)} Y_{ij} = 1 \ \forall i \in V, \ 	
	L(Y) + \alpha \one \one^T - B \succeq 0,\ B_{ij} = 0\ \forall i \neq j, Y_{ij} \geq 0\ \forall i,j \mcom 
\end{flalign*}
and that the optimal solutions to the primal and the dual must satisfy 
\begin{equation}
\sum_{i \in V} \eta_i \paren{1 - \sum_{j \in N(i)} Y_{ij}} = 0 \qquad \textrm{ and} \qquad 
	 \paren{L(Y) + \alpha \one \one^T - B} \cdot U = 0 \mper
\end{equation}

We note that the primal SDP is strictly feasible, and hence strong duality holds.

\subsection{Proof of \prettyref{prop:expansion-property}} \label{sec:expansion-property}
\begin{proof}
If $G=(V,E)$ is an undirected graph with uniform edge weights and eigenvalue gap $\lambda_2$,  the \emph{stationary distribution} is given by :

\begin{equation}
\mu_i ~=~ \frac{\Delta(i)}{ \sum_{j \in V} \Delta(j)} \geq \frac{1}{nr}  \qquad \qquad \ldots \text{ since $ \forall i, \, \Delta(i) \in [d,  rd]$} 
\end{equation}

 From the definition of $\lambda_2$, we have, for any $X \in \R^n$:

\begin{align*}
 \frac{1}{|E|}\sum_{ij \in E} (X_i - X_j)^2 &~\geq~ \lambda_2 \sum_{i,j \in V} \mu_i \mu_j (X_i - X_j)^2 \\
&~\geq~  \frac{\lambda_2}{r^2 n^2}\sum_{ij \in V} (X_i - X_j)^2
\end{align*}

Since $|E| \geq nd/2$, rearranging the above immediately yields the proof of \prettyref{prop:expansion-property}.
\end{proof}

\subsection{Proof of \prettyref{lem:flow-routing-suffices}} \label{app:flow-routing-proof}
\begin{proof}
Suppose the flows are defined by a set of paths $\mP_{it}$ between every $i \in S\setminus T$, and $t \in T$. For every $\gamma \in \mP_{it}$, let $f(\gamma)$ be the flow routed along that path. For such  a path, denote $|\gamma|$ as the length of the path, and use $(i_1=i, i_2, \ldots, i_{|\gamma|}=t)$ for the vertices along the path. For any $X \in \R^n$,  we can write:

\begin{align*}
X^T L(Y)_{\mid S}X &~\geq~ \sum_{\substack{i \in S \setminus T\\ t \in T}}\paren{ \sum_{(i_1, \ldots ,i_{|\gamma|}) := \gamma \in \mP_{it}} f(\gamma) \sum_{k \in [|\gamma|-1]} (X_{i_k} - X_{i_{k+1}})^2} \\    
&~\geq~ \sum_{\substack{i \in S \setminus T\\ t \in T}}\paren{ \sum_{ \gamma \in \mP_{it}} f(\gamma)\frac{1}{|\gamma|} (X_{i} - X_{t})^2} \qquad \qquad \ldots \text{ using \prettyref{fct:l-triang-ineq}}\\
&~\geq~ \sum_{\substack{i \in S \setminus T\\ t \in T}}(X_{i} - X_{t})^2 \paren{\frac{1}{l} \sum_{\gamma \in \mP_{it}}  f(\gamma)} \qquad \qquad  \ldots \text{ since $|\gamma| \leq l$}\\
&~\geq~ \frac{c'}{n}\sum_{\substack{i \in S \setminus T\\ t \in T}}(X_{i} - X_{t})^2 \qquad \qquad \qquad  \ldots \text{ since $\sum_{\gamma \in \mP_{it}}  f(\gamma) \geq c'l/n$}
\end{align*} 

\end{proof}


\subsection{Proof of \prettyref{lem:cluster-suffices}} \label{app:cluster-suffices-proof}

We first state and prove a lemma concerning embeddings of the graph into the real line $\R$. This is a slight variant of \cite[Lemma 9.5]{lrv13}. We include a proof here for the sake of completeness.

\begin{lemma}\label{lem:l1-embedding}
If there is a mapping $y:V \rightarrow \R$ that satisfies:
\[
\frac{ n \sum_{i} \max_{e=\{i,j\}}\abs{y_i - y_j}} {\sum_{i,j \in V} \abs{y_i - y_j}} \, = \, \delta_0 
\]
Then there is an polynomial-time algorithm to find a cut $(W, W')$ vertex expansion at most $2\delta_0$.
\end{lemma}

\begin{proof}
Let us assume without loss of generality that $y_1 \leq y_2 \leq \ldots \leq y_n$. Let $S_k$ denote the level cut $\{1,2,\ldots k \}$, for $1 \leq k < n$. The algorithm simply outputs the cut among the $S_k$'s that has the minimum  (balanced) vertex expansion. Let us define $\alpha_k \eqdef \abs{y_{k+1} - y_{k}}$. We have:

\begin{align}
\min_k \phiv(S_k) & ~=~ \min_{k} \phiv(S_k) \\ 
&~=~ \min_{k} \frac{n \sum_{i \in V} \one_{\insquare{i \text{ connects across } S_k}}}{\sum_{i,j} \one_{\insquare{i,j \text{ are on opposite sides of } S_k}}} \\
&~\leq~ \frac{n \cdot \sum_k  \alpha_k  \sum_{i \in V} \one_{\insquare{i \text{ connects across } S}}} {\sum_k \alpha_k \cdot \sum_{i,j} \one_{\insquare{i,j \text{ are on opposite sides of } S_k}}} \\  
&~=~ \frac{n \cdot \sum_i \paren{ \sum_{k : i \in \text{bdry} (S_k) } \alpha_k} }{\sum_{i,j} \sum_k \one_{\insquare{S_k \text{ separates } i,j}} \cdot \alpha_k} \label{eq:l1-eqn-1}
\end{align}

The inequality above follows from the following fact: if $a_1, a_2, \ldots ,a_n, b_1, \ldots, b_n$ are non-negative reals, then: $\min_{i \in [n]} \frac{a_i}{b_i} \leq \frac{\sum_i a_i}{\sum_i b_i}$.

Let us first consider the numerator in \prettyref{eq:l1-eqn-1}. For any fixed $i \in V$, let $j^*(i) \eqdef \arg \max_{j \in N(i)} |y_i - y_j|$.  Suppose $j^*(i) \geq i$ without loss of generality; the cuts $S_k$ that put $i$ on the boundary for $k \geq i$ are precisely the cuts $S_{i}, \ldots S_{j-1}$.  Thus, we have: 

\[
\sum_{k\geq i : i \in \text{bdry} (S_k)}  \alpha_k ~=~ 
\sum_{k=i}^{j^*(i)-1}  \alpha_k ~=~ y_{j^*(i)} - y_i 
\]

The contribution from cuts $S_k$ for $k <i$ is also bounded by  the above quantity. Hence, we have that the numerator is upper bounded by:
\[
\sum_{k : i \in \text{bdry} (S_k)}  \alpha_k \leq 
2 \cdot \vert y_{j^*(i)} - y_i \vert 
\]

\smallskip

It remains to lower bound the denominator. This is easy, as the following identity is immediate from the definition of $\alpha_k$, for any fixed $i,j$ (again, without loss of generality, let $j > i$):

\[
\sum_{i,j} \sum_k \one_{\insquare{S_k \text{ separates } i,j}} \cdot \alpha_k = \sum_{k=i}^{j-1} \alpha_k ~=~ |y_j - y_i|
\]

Combining the two expressions, we get that:
\[
\min_k \phiv(S_k) ~\leq~ 2 \cdot \frac{ n \sum_{i} \max_{e=\{i,j\}}\abs{y_i - y_j}} {\sum_{i,j \in V} \abs{y_i - y_j}} \, \leq  \, 2\delta_0
\]

It is easy to see that the best $S_k$ can be found efficiently, given the $y_i$'s.
\end{proof}

\medskip
We are now ready to give the proof of \prettyref{lem:cluster-suffices}. We introduce some notation for clarity: for $i,j \in V$, denote $d(i,j)= \norm{u_i - u_j}^2$.  Since the $u_i$'s obey $\ell_2^2$ triangle inequalities, $d(\cdot, \cdot)$ is a distance function. In a natural fashion, for any $L \subseteq V$, we will denote $d(i, L) \eqdef  \min_{j \in L} \norm{u_i - u_j}^2$. We refer to the squared distances measured by $d(\cdot, \cdot)$ as $\ell_2^2$ distances.

\medskip
\begin{proof}[Proof (Of \prettyref{lem:cluster-suffices})]

We are given that there exits a set $L$ such that $\abs{L} \geq \alpha n$, satisfying:

\[
 \E_{i,j \in L } \norm{u_i - u_j}^2 \leq \frac{1}{40}
\]

Thus, there exists an $i_0 \in L$, with $\E_{j \in L} [ \norm{u_{i_0} - u_j}^2] \leq 1/40$ . Since $\abs{L} \geq \alpha n$, by Markov's inequality, an $\ell_2^2$ ball of radius $1/10$  around $i_0$ should have:

\[
\abs{B(i_0, \frac{1}{10})} \geq \frac{ 3\alpha n}{4}
\]

We will set $L' \eqdef B(i_0, 1/10)$, thus $\abs{L'} \geq 3 \alpha n/4$. 
%

\begin{Claim}
For any $i \in V$, we have $\abs{B(i, 1/8)} \leq 14n/15$
\end{Claim}

\begin{proof}
Suppose $\abs{B(i, 1/8)} = k$. Then, since $\sum_{i<j} d(i,j) = n^2$, we should have:
\begin{align*}
2n^2 \leq k^2 \cdot \frac{1}{4} + (n^2 - k^2)\cdot 4 \\
\implies k^2 \leq \frac{2}{4-\frac{1}{4}} n^2  \leq \frac{2}{3} n^2~\implies k \leq \frac{14}{15}n
\end{align*}

The first inequality is true because two points within  an $\ell_2^2$ distance $1/8$  of $i$ are at most $1/4$ apart in squared distance, since they obey $\ell_2^2$ triangle inequalities. Also the points, being unit vectors, are all within a ball of $\ell_2^2$ diameter $4$. 
\end{proof}

Let $R' \eqdef V \setminus B(i_0, 1/8)$. Note that $L' \subseteq B(i_0, 1/8)$ and is disjoint from $R'$.  Furthermore, $d(L', R') \geq 1/50$. Now, consider the mapping $y:V \rightarrow R^{+}$:

\[
y_i \eqdef 
\begin{cases}
d(i, L'), & \text{for } i \notin R'\\
d(R', L') & \text{for } i \in R' \mper
\end{cases}
\]

We show that the mapping $y_i$  satisfies the conditions of \prettyref{lem:l1-embedding}, with $\delta_0 = O_{c_0, c_1}(\delta)$.
\smallskip

The following fact is an easy consequence of the $\ell_2^2$ triangle inequality.

\begin{fact} \label{fct:setdistance}
Let $V$ be a set of points satisfying $\ell_2^2$ triangle inequalities, and $L \subseteq V$. Then for any $i,j \in V$, we have $\abs{d(i,L) - d(j,L)} \leq d(i,j)$.
\end{fact}
\begin{proof}
Let $d(j, L) \geq d(i,L)$ without loss of generality. Let $i' \in L$ be such that $d(i,L)= d(i,i')$. By definition, $d(j, i') \geq d(j,L)$. This gives us
\[ \abs{d(i,L) - d(j,L)} = d(j,L) - d(i,L) \leq d(j, i') - d(i,i') \leq d(j,i) \mcom
\]
where the last inequality used the $\ell_2^2$ triangle inequality.
\end{proof}

For any fixed $i$, we have that:
\begin{align*}
\abs{y_i - y_j} &\leq  \abs{d(i,L') - d(j,L')}\\
&\leq  d(i,j) \qquad \qquad \ldots \text { Using \prettyref{fct:setdistance}}
\end{align*}
Note that the first inequality holds even when one of the two points, say $j$, is in $R'$, as the points from $R'$ are the furthest in the line embedding from $L'$, as compared to other points in $V$, and hence $d(j, L') \geq d(R', L')=y_j \geq y_i$ for every $j \in R', i \notin R'$. Thus, we have:

\[
\max_{j\in N(i)}  \abs{y_i - y_j} \leq \max_{j\in N(i)} d(i,j)
\]
\smallskip

Next, we analyze the following sum:

\begin{align*}
\sum_{ij \in V} \abs{y_i - y_j} &\geq  \sum_{\substack{i \in L'\\ j \in R'}}  \abs{d(i,L') - d(R',L')} \\
& = \abs{L'} \abs{R'}d(R', L') \\
& \geq \Omega(\alpha) \cdot n^2 
~ \geq \Omega(\alpha) \sum_{i,j} \norm{u_i - u_j}^2 \qquad \ldots \text{ from the SDP balance constraint.}
\end{align*}

Combining the above, we get that:
\[
\frac{ n \sum_{i} \max_{e=\{i,j\}}\abs{y_i - y_j}} {\sum_{i,j \in V} \abs{y_i - y_j}} \, \leq  \, O(1) \cdot \frac{ n \sum_{i \in V} \max_{j \in N(i)} \norm{u_i - u_j}^2}{\alpha \sum_{ij} \norm{u_i - u_j}^2}
 ~\leq~ O(\delta/\alpha) 
\]

Using \prettyref{lem:l1-embedding}, we conclude that we can find a patrition $(W, W')$ such that the number of boundary vertices is $O(\delta/\alpha)n$. Due to the nature of our embedding, $W$ satisfies $B(i, 1/10) \subseteq W \subseteq B(i, 1/8)$, so $|W| \in [ \Omega(\alpha n),  14n/15]$.  We state the algorithm in Algorithm 1. The algorithm explicitly searches for the correct set $W$.
\end{proof}


\section{Proof of $O(\sqrt{\epsilon})$ guarantee for \cite{lr14}'s model} \label{app:lr14}

\begin{definition}[Planted model, \cite{lr14}]

An instance from the planted model $G\sim$ \planted$(n,\eps, \lambda)$ is generated as follows: 

\begin{enumerate}
\item Partition the vertex set of size $n$ into two equal halves $(S, S')$ arbitrarily.

\item Within $S$, and $S'$, the add edges so that $G[S]$ and $G[S']$ are regular edge expanders with spectral gap $\lambda$.

\item Choose sets $T \subseteq S$ and $T' \subseteq S'$ each of size $\epsilon n$ arbitrarily, and add edges between them.

\item (Monotone Adversary) Add arbitrary edges within $S$ and $S'$.
\end{enumerate}

Output the generated instance $G$.

\end{definition}
\medskip

\begin{algorithm} \label{alg:vexp-alg-1}
\caption{Algorithm for Vertex Expansion in the planted model}
\begin{algorithmic}[1]
\INPUT $G=(V,E)$, and an optimal SDP solution $\inbrace{u_i}_{i\in V}$ with value $\delta n$.
\OUTPUT A partition $(W^*, V\setminus W^*)$ of $V$ with both parts of size $\Omega(n)$.

\State Let $B \eqdef \set{i\in V ~:~\eta_i \geq \sqrt{\delta}}$. $V'  \gets V\setminus B$ .   
 \For{$t = 50, \ldots , \ceil{\frac{1}{50\sqrt \delta}}+1$}
    \State $W^*_t \; \gets \; \arg\min_{W \in \mathcal{W}} \,\,\,  \phiv(W)\quad \text{ where } \mathcal{W}=\set{B_{V'}(i, t\sqrt{\delta})} {i \in V'} $.
\EndFor
\State Output $W^* \eqdef \arg\min_t \phiv(W^*_t \cup B)$.
\end{algorithmic} 
\end{algorithm}

\begin{theorem}\label{thm:vexp-alg-1}
Let $G$ be an instance from \planted$(n,\eps, \lambda)$, where $\eps$ is small enough. Algorithm 2 outputs a set $W^*$ of size $\Omega(n)$ such that $\phiv(W^*) \leq O(\sqrt\eps)$.
\end{theorem}
\begin{proof}
 Let $\phisdp(G) =\delta \leq 8\epsilon n$. We denote the SDP solution by vectors $u_i \in \R^n$, for $i \in V$, which satisfy satisfy $u_i^T u_j = U_{ij}$ . In the optimal SDP solution $(U, \eta)$ observe that  $\eta_i = \max_{j \in N(i)} \norm{u_i- u_j}^2$. Define $B_S(i,r) \defeq \set{j\in S : \norm{u_i- u_j}^2 \leq r}$ for what follows. We will assume that $\epsilon$ is small enough.

Clearly, the set $B$ in step 1 of the algorithm satisfies $|B| \leq \sqrt{\eps} n$, using Markov's inequality. Since the vectors are all on the unit ball, we have $\norm{u_i-u_j}^2 \leq 2$. As we have discarded at most $\sqrt{\epsilon} n^2$ pairs of vertices by removing $B$, the set $V'$ satisfies:

\[
    \E_{ij \in V'} \insquare{\norm{u_i- u_j}^2} \geq {\frac{1}{4}-2\sqrt \eps}. 
\]

\medskip
\begin{proposition}\label{prop:max-points-in-ball}
For any $u\in V'$, $\abs{B_{V'}(u,1/50)} \leq 9n/10$. 
\end{proposition}
\begin{proof}
Suppose not. Then the average distance in $V'$ would be at most $ \frac{1}{10}.2 + (\frac{9}{10})^2 \cdot \frac{1}{25} < 1/4 -2\sqrt{\eps}$, a contradiction.
\end{proof}

Let $S, S'$ be the original planted cut in the graph. Define $\tilde{S} \eqdef { S\setminus B}$, and $\tilde{S}' \eqdef V' \setminus S$.

\begin{observation}\label{obs:avg-corr-S}
With high probability (over the choice of the instance $G$), we have:

\begin{equation}\label{eq:avg-pair-in-component}
\E_{ij\in \tilde{S}} \insquare{\norm{u_i- u_j}^2} \leq O(\sqrt{\epsilon}), \qquad \text{and} \qquad \E_{ij\in \tilde{S}'} \insquare{\norm{u_i- u_j}^2} \leq O(\sqrt{\epsilon}) 
\end{equation}
\end{observation}
\begin{proof}
Consider $\tilde{S}$. Since we have discarded \emph{all} long vertices $B$, we have $\eta_i \leq \sqrt{\epsilon}$, for each $u\in \tilde{S}$. Since $G[\tilde{S}]$ is a random graph, it is a spectral expander (with constant spectral gap) with high probability. This yields:

\begin{equation}
\E_{ij \in \tilde{S}}\insquare{\norm{u_i- u_j}^2} ~{\leq}~ O(1) ~ \E_{i, j \in E[ \tilde{S}]}\insquare{\norm{u_i- u_j}^2} ~{\leq}~ O(1)~\max_{i \in \tilde{S}} \, \eta_i \leq O(\sqrt{\eps}). 
\end{equation}

A similar argument holds for $\tilde{S}'$.
\end{proof}

Thus, there exists a $i\in \tilde{S}$ such that  $\E_{j \in \tilde{S}} \insquare{\norm{u_i- u_j}^2} \leq O(\sqrt{\epsilon}) \leq c\sqrt{\epsilon}$, for some constant $c$.  By Markov's inequality, there are at least $n/20$ vertices in the $\ell_2^2$ ball of radius $50c \sqrt{\epsilon}$ around $i$ (since $|\tilde{S}| \geq n(\frac{1}{2} -\sqrt{\eps}) \geq n/4$). We now can claim the following:

\begin{Claim}\label{claim:app-claim}
There is a constant $t \in \N$, $t\leq \frac{1}{50\sqrt\eps}$ such that $|B_{V'}(i, (t+1) \sqrt{\eps})| \leq (1+\sqrt \eps)| B_{V'}(i, t\sqrt{\eps})|$
\end{Claim}
\begin{proof}
Let $t_0 \eqdef \ceil{50c} $. Fix any $t>t_0$, and suppose  the claim is not true for any $t' \leq t$. Then:
\begin{align*}
\frac{9n}{10} &\geq |B_{V'}(i, t \sqrt{\eps})|\\ 
&\geq (1+\sqrt\eps)^{t-t_0} |B_{V'}(i, t_0 \sqrt{\eps})|\\
&\geq (1+\sqrt \eps)^{t-t_0} n/20 
\end{align*}

The first inequality follows from Proposition~\ref{prop:max-points-in-ball}. 
This implies $t \leq t_0 + \log_{1+\sqrt{\eps}}(18)$, which is less than $1/50\sqrt{\eps}$, for a small enough $\eps$.\end{proof}

Thus, for the appropriate setting of $t=t^*$,  the set of vertices in the ball $B_{V'}(i, t^*\sqrt{\eps})$ has at most $\sqrt{\eps}n$ points on the boundary of the cut (in $G$). This is because all of these vertices satisfy $\eta_i \leq \sqrt{\eps}$ and hence their neighbors outside the set should lie in $B(i, (t^*+1)\sqrt{\eps})$.  These are bounded in number by $|B_{V'}(i, (t^*+1) \sqrt{\eps}) \setminus B_{V'}(i, t^* \sqrt{\eps})|$. From \prettyref{claim:app-claim}, this is at most $\sqrt{\eps}n$ for our choice of $t^*$. 

Furthermore, adding $B$ of size at most $\sqrt \eps n$ to this increases the number of boundary points on the cut by at most $|B| = O(\sqrt{\eps} n)$.

Finally, the theorem follows, by observing that this ball around $i$ is considered as a candidate $W$ in step 3 of Algorithm~1.

\end{proof}


\section{Instances where using edge-expansion does not suffice}\label{app:edgeexpansionbad}

Our model for $\vbm$ allows arbitrary addition of edges in a monotone manner anywhere in the graph. This contrasts with known monotone models for edge expansion, where edges across the planted cut can only be deleted or reduced. To the best of our knowledge, in its full generality, $\vbm$ captures instances that do not fall under any of the block models studied before. To make this explicit, we give below a family of instances $\mathcal{H}$ that are not captured by the edge-expansion block models considered in literature (see \prettyref{sec:relatedwork}). 

\smallskip
\begin{definition}(Bad instances $H_n$ for edge expansion)
$H_n$, a class of graphs on $n$ vertices occurring in $\vbm$, is generated in the following way: We fix $\eps = \omega(n^{-1/3})$ and $p = \log n/n$. We let $\eps_1 = \eps_2 = \eps$, so the sets $T, T'$ are both of size $\eps n$. Arbitrarily partition $S\setminus T$ into two parts: $A_1, A_2$, with $\abs{A_i} = \abs{S\setminus T}/2 = n(1-\epsilon)/4$ and $A_1 \cup A_2= S \setminus T$. The corresponding sets in $S' \setminus T'$  are called $B_1$ and $B_2$. Similarly, partition $T$ into $T_1, T_2$, each of size $\abs{T}/2$, and $T'$ into $T'_1, T'_2$.

\smallskip
We now describe the edges in the graph. We first set $S\setminus T$ (and $S' \setminus T'$) to be an arbitrary, constant degree spectral expander, with eigenvalue gap $\lambda$. On $T_i \times T'_i$, for $i=1,2$, we include all edges, making them complete bipartite graphs. On the pairs in $S \setminus T \times  T$ and $S' \setminus T' \times  T'$, we generate the edges randomly and independently with probability $p$. 

\smallskip
Over the edges present in the base graph described above, add edges arbitrarily within pairs $A_i \times A_i$, $B_i \times B_i$, $A_i \times T_i$ and $B_i \times T'_i$ so that the degrees of all nodes in the graph are $\alpha n \pm o(n)$, for some small constant $\alpha \in (0,1/2)$. 
\end{definition}

\begin{figure}
\centering
\label{fig:edge-expansion-bad}
\includegraphics[scale=1.0]{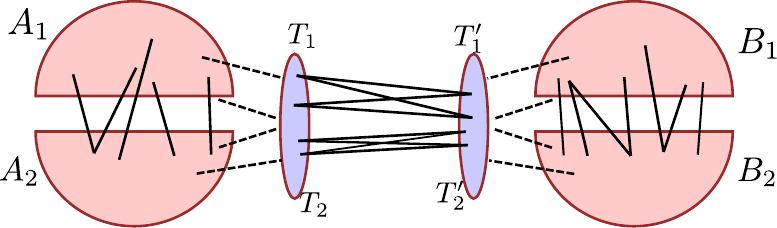}
	\caption{Structure of instances where edge expansion  based algorithms fail.}
\end{figure}

\begin{lemma}\label{lem:edge-exp-bad-lemma}
For the family of graphs $H_n$, the following holds with high probability:
\begin{enumerate}
\item There is a balanced cut in the graph, with $O(n)$ edges going across it, corresponding to $A_1 \cup B_1 \cup T_1 \cup T'_1$ and $A_2 \cup B_2 \cup T_2 \cup T'_2$.
 
\item Any cut other than $(S, S')$ in the graph has $\Omega(n)$ vertices on its vertex boundary. Thus, the sparsest balanced vertex cut is $(S, S')$, and it has $\Omega(\epsilon^2 n^2) = \omega(n^{4/3})$ edges going across.


\end{enumerate}
\end{lemma}

\begin{proof}
$(1)$ follows almost immediately from the definition of $H_n$, by noting that the randomness adds at most $o(n)$ edges across the stated cut, with high probability.

For $(2)$, consider any cut that cuts $S$ into two parts $R, R'$ (for $S'$, a similar argument will apply). Without loss of generality, suppose $|R \cap T| \geq |T|/2$.  We have the following cases:
\begin{enumerate}
\item $\abs{R \cap (S \setminus T)} \leq \alpha n/2$. Then some vertex in $R \cap T$ has $\alpha n/2$ neighbors in $R'$, giving $\Omega(n)$ vertices on the boundary.

\item  $(1-\alpha)n/2 \geq \abs{R \cap (S \setminus T)} \geq \alpha n/2$: Consider the subgraph on $S \setminus T$; since $\abs{R' \cap (S\setminus T)} \geq \alpha n/2$, and $G[S\setminus T]$ contains a constant-degree spectral expander, we have that the cut $R , R'$ restricted to $S \setminus T$ should have $\Omega(n^2)$ edges going across it. Consequently, the vertex boundary of the cut is of size $\Omega(n)$.

\item $\abs{R \cap (S \setminus T)} \geq (1-\alpha)n/2$. This means that $\abs{R' \cap (S \setminus T)} \leq \alpha n/2$. This implies that every vertex in $R$ has at least $\alpha n/2 -o(n)$ neighbors outside $R'$, since they have degrees $\alpha n \pm o(n)$, and hence the vertex boundary is again $\Omega(n)$.
\end{enumerate}

\end{proof}

Given the above, since edge-expansion based algorithms only recover almost-balanced cuts with $O(n)$ edges (even in the approximate setting), their sets will have an asymptotically larger number ($\Omega(n)$) of vertices on the boundary of the cut.  However, \prettyref{thm:sdpintegral} shows that the vertex-expansion SDP  exactly recovers the intended cut $S, S'$, which has only $o(n)$ vertices on its boundary.

\medskip

Our construction for the model for vertex-expansion, and the above graphs in particular can be motivated by the notion of \emph{hubs} and resulting structure of graphs occurring in real-life networks. The sets $S\setminus T$ with connections of the above form can be thought of as local communities connecting to each other loosely,  and to \emph{hubs}, represented by the vertices in $T$ and $T'$. The hubs themselves are a small set of vertices having dense connections between each other, and are the `critical' vertices in the graph. In order to detect the hubs in such a setting, edge-expansion block models seem to fall short.

\newcommand{\expo}[1]{e^{#1}}
\newcommand{\vol} {\mathrm{vol}}
\section{Bipartite Graphs and Expansion} \label{app:bipartite-expansion-section}

We restate \prettyref{prop:br-1} for clarity, and give its proof.

\begin{proposition}[(\prettyref{prop:br-1} restated)]
\label{prop:br} 
Let $L,R$ be two (disjoint) sets of vertices of sizes $(1 - \gamma) n$ and $\gamma n$ respectively, where $\gamma \leq 1/2$.
Let $p \in (0,1)$ be any number satisfying $p \gamma n \geq 300 \log n$.	
Let $\tilde{G}$ be a random bipartite graph obtained by adding edges between each pair in $L \times R$
independently with probability $p$. 
Let $u_1, \ldots, u_n$ be a set of vectors. Then, there exists an absolute constant $C$ such that, with high probability, we have:
\begin{enumerate}
 \item[a)] \[ \E_{i,j \in L} \norm{u_i - u_j}^2 \leq C \E_{\set{ij} \in E(\tilde{G})} \norm{u_i - u_j}^2  \mper  \]
\item[b)]The  minimum and maximum degrees $\Delta_{\min}(L),~\Delta_{\max}(L)$ in $L$ satisfy $\Delta_{\max}(L) / \Delta_{\min}(L)  \leq 3$.
\end{enumerate}

\end{proposition}

\begin{lemma}
\label{lem:br-deg}
The graph $G$ satisfies  
\[ \frac{5}{6} \Ex{\Delta(i)} \leq \Delta(i) \leq \frac{7}{6} \Ex{\Delta(i)} \qquad \forall i \in L \cup R \mper \]
with probability at least $1 - 1/n^2$. 
\end{lemma}

\begin{proof}
This lemma follows by a straight forward application of the Chernoff bound. 

Fix a vertex $i \in L$. Then, $\Ex{\Delta(i)} = p \gamma n$. Using the Chernoff bound, 
\[ \Pr{ \frac{5}{6} p \gamma n \leq \Delta(i) \leq \frac{7}{6} p \gamma n}	
	\geq 1 - 2e^{p \gamma n/72} \geq 1 - \frac{2}{n^4} \mper \]
Fix a vertex $j \in R$. Then, $\Ex{\Delta_j} = p (1 - \gamma) n$. Using the Chernoff bound, 
\[ \Pr{ \frac{5}{6} p (1 - \gamma) n \leq \Delta(i) \leq \frac{7}{6} p (1 - \gamma) n}	
	\geq 1 - 2e^{p (1- \gamma) n/72} \geq 1 - \frac{2}{n^4} \mper \]
Using a union bound over all the vertices $i \in L \cup R$, we get that 
\[ \Pr{ \frac{5}{6} \Ex{\Delta(i)} \leq \Delta(i) \leq \frac{7}{6} \Ex{\Delta(i)} \ \forall i \in L \cup R} \geq 1 - \frac{1}{n^2} \mper \]

\end{proof}

Next, we show that the number of edges crossing any set is close to its expected value.

\begin{lemma}
\label{lem:br-edges}
For this graph $G$, we have 
	\[ \Abs{E(U, U')} \geq \frac{1}{2} \Ex{\Abs{E(U, U')}} \qquad \forall U \subset L \cup R \]
with probability at least $1 - 2/n^2$.

\end{lemma}

\begin{proof}
We will prove this lemma by using the Chernoff bound to bound the number of edges crossing a fixed
subset $U$, followed by a union bound over all the subsets.

\medskip

Fix any non-empty set $U \subset L \cup R$. Let $U'$ denote $(L \cup R) \setminus U$.
Let $a = \Abs{U \cap L}$ and $b = \Abs{U \cap R}$. Without loss of generality, we can assume that $a+b \leq n/2$, or else, we could run the argument on $U'$ instead.
Then,
\begin{equation}
\label{eq:br-edges-1}
\Ex{\Abs{E(U, U')}} = p \paren{a (\Abs{R} - b) + b (\Abs{L} - a)}  \mper 
\end{equation}
Therefore, using the Chernoff bound, 
\begin{align*} 
	\Pr{ \Abs{E(U, U')} \leq \frac{1}{2} \Ex{\Abs{E(U, U')}}}
	& \leq \expo{- p \paren{a (\Abs{R} - b) + b (\Abs{L} - a)}/8} 
	= \expo{- p \paren{a (\gamma n - b) + b ((1 - \gamma)n - a)}/8} 
\end{align*}

For convenience, we define the function  $f(a,b)$ and note a useful lower bound on it below:
\begin{align*}
f(a,b) &\defeq \frac{1}{8} p \paren{a (\gamma n -  b) + b ((1-\gamma)n-a)}\\
&= \frac{1}{8} p \paren{a (\gamma n -  2b) + b (1-\gamma)n} \\
& \geq \frac{1}{8} p \paren{a (\gamma n -  2b) + \frac{bn}{2}} & \ldots \text{ using }  \gamma \leq \frac{1}{2}
\end{align*}
Using the union bound over all subsets $U$, we get  
\begin{multline} 
	\Pr{ \Abs{E(U, U')} \geq \frac{1}{2} \Ex{\Abs{E(U, U')}} \ \forall U \subset L \cup R 
		\textrm{ such that }  \abs{U} \leq n/2 }\\
	 \geq 1- \sum_{b = 0}^{\gamma n} \sum_{a = 0}^{(1 - \gamma)n} 
	 \binom{\gamma n}{b} \binom{(1 - \gamma)n}{a} \expo{-f(a,b)} \mper 
\end{multline}
We analyze this expression in the following cases.
\begin{enumerate}
\item[Case 1:] {\em $b \leq \gamma n/4$.}
In this case, we have:
\[ f(a,b) ~\geq ~\frac{1}{8} p \paren{a (\gamma n - 2 b) + b n/2} 
\geq \frac{1}{16} a p \gamma n + \frac{1}{16} b p n \geq 5 (a + b) \log n \mper  \]
Above, the first inequality holds because $\gamma \leq 1/4$, the second by the bound on $b$, and the third by the bound on $p\gamma n$. Therefore, 
\begin{align}
\sum_{b = 0}^{\gamma n/4} \sum_{a = 0}^{(1 - \gamma)n} \binom{\gamma n}{b} \binom{(1 - \gamma)n}{a} f(a,b)
	& \leq  \sum_{b = 0}^{n} \sum_{a = 0}^{n} n^a n^b e^{- 5 (a + b) \log n}  \nonumber \\
	& = \sum_{b = 0}^{n} \sum_{a = 0}^{n} \frac{1}{n^{4(a+b)}}
	\leq  \sum_{b = 0}^{n} \sum_{a = 0}^{n} \frac{1}{n^4} & \paren{\textrm{Since } a + b \geq 1} \nonumber \\
	& \leq \frac{1}{n^2}  \mper \label{eq:br-edges-2} 
\end{align}
\item[Case 2:] {\em $\gamma n /2 \geq b > \gamma n / 4$.}
In this case, 
\[ f(a,b) \geq \frac{1}{8} p \paren{a (\gamma n - 2 b) + b n/2} \geq \frac{1}{64} p \gamma n^2 \geq 4 n \log n \mcom \]
 where the second inequality follows by ignoring the first term $a (\gamma n - 2 b)$, which is non-negative given the upper bound on $b$. Therefore, 
\begin{equation}
\label{eq:br-edges-3}
\sum_{b = \gamma n/4}^{\gamma n/2} \sum_{a = 0}^{(1 - \gamma)n} \binom{\gamma n}{b} \binom{(1 - \gamma)n}{a} \expo{-f(a,b)}
	 \leq \sum_{b = 0}^{\gamma n} \sum_{a = 0}^{(1 - \gamma)n} \binom{\gamma n}{b} \binom{(1 - \gamma)n}{a}
	e^{-4 n \log n} = 2^n e^{-4 n \log n} \mper 
\end{equation}

\item[Case 3:] {\em $b \geq \gamma n /2 $.}  Since $a+b \leq n/2$, we have $a \leq n (1-\gamma)/2$. We start with the definition of $f(a,b)$ and lower bound it. First, note that if $a=0$, then 
\begin{align*}
f(a,b) ~\geq~ \frac{1}{8}pb(1-\gamma)n 
~\geq~ \frac{1}{32} p \gamma n^2 
 ~\geq~ 8n \log n \mper
\end{align*}

The second inequality used the lower bound on $b$ and that $\gamma \leq \frac{1}{2}$. If $a \geq 1$, then:
\begin{align*}
f(a,b) &=   \frac{1}{8} p \paren{a (\gamma n - b) + b ((1- \gamma)n-a) } \\
& = \frac{1}{8} p \paren{a \gamma n   + b (1-\gamma)n -2ab} \\
& \geq  \frac{1}{8} p \paren{a \gamma n   + b (1-\gamma)n -2b (1-\gamma) \frac{n}{2}} & \ldots \text{since } a \leq \frac{(1-\gamma)n}{2} \\
& \geq \frac{1}{8}a p \gamma n \\
& \geq \frac{1}{8} p \gamma n \geq 32n \log n 
\end{align*} 

Hence, similar to the previous case, we get:
\begin{equation}
\label{eq:br-edges-4}
\sum_{b = \gamma n/2}^{\gamma n} \sum_{a = 0}^{(1 - \gamma)n} \binom{\gamma n}{b} \binom{(1 - \gamma)n}{a} \expo{-f(a,b)}
\leq 2^n e^{-8 n \log n} \mper 
\end{equation}

\end{enumerate}
Using \prettyref{eq:br-edges-2}, \prettyref{eq:br-edges-3} and \prettyref{eq:br-edges-4},  we get that
\[
\Pr{ \Abs{E(U, U')} \geq \frac{1}{2} \Ex{\Abs{E(U, U')}} \ \forall U \subset L \cup R 
		\textrm{ such that }  |U| \leq n/2 }
	 \geq 1 - \frac{2}{n^2} \mper \]

The statement of the lemma follows, since for any $U$, either $\abs{U} \leq n/2$, or $\abs{U'} \leq n/2$.
\end{proof}

Next, we show that $G$ is an edge expander.
\begin{lemma}
\label{lem:br-eig}
$G$ satisfies the following properties with high probability.
\begin{enumerate}
\item $\paren{\max_{i \in L} \Delta(i)} / \paren{\min_{i \in L} \Delta(i)} \leq 2$.
\item The spectral gap of $G$ satisfies $\lambda_G \geq \lambda$, for some absolute constant $\lambda$.
\end{enumerate}	
\end{lemma}

\begin{proof}
Using a union bound, we get that w.h.p., $G$ satisfies the guarantees 
of both \prettyref{lem:br-deg} and \prettyref{lem:br-edges}. For the rest of the proof, we will
assume that $G$ satisfies these guarantees.

\smallskip
Using \prettyref{lem:br-deg}, $(5/6)\Ex{\Delta(i)} \leq \Delta(i) \leq (7/6) \Ex{\Delta(i)} $
$\forall i \in L$ with high probability. Therefore, 
$$\frac{{\max_{i \in L} \Delta(i)} }{{\min_{i \in L} \Delta(i)}} \leq \frac{7}{5} \leq 2 \mper$$

\medskip
We now bound the edge expansion of $G$. 
Fix any non-empty set $U \subset L \cup R$.
Let $U'$ denote $(L \cup R) \setminus U$.
Let $a = \Abs{U \cap L}$ and $b = \Abs{U \cap R}$.
Then, 
\begin{align}
\phi(U)  & =  \frac{ \Abs{E(U, U')} }{ \min \set{ \sum_{i \in U} \Delta(i), \sum_{i \in U'} \Delta(i)}} & \ldots \text{By defintion of $\phi$}\nonumber \\
	& \geq \frac{ \Ex{\Abs{E(U, U')}}/2 }{ \min \set{ \sum_{i \in U} (7/6) \Ex{\Delta(i)}, 
		\sum_{i \in U'} (7/6) \Ex{\Delta(i)}} } 
		& \paren{\textrm{Using \prettyref{lem:br-edges} and \prettyref{lem:br-deg} }} \nonumber \\
	& = \frac{1}{3} \frac{a (\Abs{R} - b) + b (\Abs{L} - a)}{\min \set{a \Abs{R} + b \Abs{L}, 
		(\Abs{L} - a) \Abs{R} + (\Abs{R} - b) \Abs{L} }} 
		& \paren{\textrm{Using \prettyref{eq:br-edges-1}}} \label{eq:br-exp-1} 
\end{align}	
Without loss of generality, we assume that 
$a \Abs{R} + b \Abs{L} \leq (\Abs{L} - a) \Abs{R} + (\Abs{R} - b) \Abs{L} $, or equivalently,
\begin{equation}
\label{eq:br-exp-2}
a \Abs{R} + b \Abs{L} \leq \Abs{L} \Abs{R} \mper 
\end{equation}

To lower bound the value in \prettyref{eq:br-exp-1}, we consider the following cases.
\begin{enumerate}
\item[Case 1:]
{\em $a \leq \Abs{L}/2$ and $b \leq \Abs{R}/2$.}
In this case,
\begin{equation}
\label{eq:br-exp-3}
\frac{a (\Abs{R} - b) + b (\Abs{L} - a)}{a \Abs{R} + b \Abs{L}}
	\geq \frac{a \Abs{R}/2 + b \Abs{L}/2}{a \Abs{R} + b \Abs{L}} = \frac{1}{2} \mper
\end{equation}

\item[Case 2:]
{\em $a > \Abs{L}/2$ and $b \leq \Abs{R}/2$.}
In this case, using \prettyref{eq:br-exp-2}
\begin{equation}
\label{eq:br-exp-4}
\frac{a (\Abs{R} - b) + b (\Abs{L} - a)}{a \Abs{R} + b \Abs{L}}
	\geq \frac{a (\Abs{R} - b)}{a \Abs{R} + b \Abs{L}}
	\geq \frac{\Abs{L}/2 \Abs{R}/2 }{\Abs{L} \Abs{R}} = \frac{1}{4} \mper
\end{equation}

\item[Case 3:]
{\em $a \leq \Abs{L}/2$ and $b > \Abs{R}/2$.}
In this case, using \prettyref{eq:br-exp-2}
\begin{equation}
\label{eq:br-exp-5}
\frac{a (\Abs{R} - b) + b (\Abs{L} - a)}{a \Abs{R} + b \Abs{L}}
	\geq \frac{b (\Abs{L} - a)}{a \Abs{R} + b \Abs{L}}
	\geq \frac{\Abs{L}/2 \Abs{R}/2 }{\Abs{L} \Abs{R}} = \frac{1}{4} \mper
\end{equation}

\item[Case 4:]
{\em $a > \Abs{L}/2$ and $b > \Abs{R}/2$.}
In this case,
\[ a \Abs{R} + b \Abs{L} >  \Abs{L} \Abs{R}/2 +  \Abs{L} \Abs{R}/2 = \Abs{L} \Abs{R} \]
which contradicts \prettyref{eq:br-exp-2}. Therefore, this case can not arise. 

\end{enumerate}
Using \prettyref{eq:br-exp-1}, \prettyref{eq:br-exp-3}, \prettyref{eq:br-exp-4} and \prettyref{eq:br-exp-5},
we get that $ \phi(U) \geq \frac{1}{12}$.
Therefore, using Cheeger's inequality, we get that 
	\[ \lambda_G \geq \frac{1}{2} \paren{\min_{U \subset L \cup R} \phi(U) }^2 \geq \frac{1}{2\cdot 12^2} \mper  \]
\end{proof}

\begin{proof}[Proof of \prettyref{prop:br}]
Let $\mu_G$ be the stationary distribution on the graph vertices, and let $\mu_L$ denote the marginal of $\mu_G$ on $L$.

\begin{align*}
\E_{\set{ij} \in E(G)} \norm{u_i - u_j}^2 & \geq \lambda \E_{i,j \sim \mu_G} \norm{u_i - u_j}^2 & \paren{\textrm{Using \prettyref{lem:br-eig}}} \\
    & \geq \lambda \cdot \frac{1}{\paren{\sum_{i \in L\cup R} \Delta(i)}^2} \sum_{ i \in L, j \in L} \Delta(i) \Delta(j) \norm{u_i - u_j}^2 &  \paren{\textrm{From definition of $\mu_G$}} \\
	& \geq \lambda\cdot \frac{1}{4} \cdot \frac{1}{\paren{\sum_{i \in L} \Delta(i)}^2} \sum_{ i \in L, j \in L} \Delta(i) \Delta(j) \norm{u_i - u_j}^2 & \paren{ \textrm{Since } \mu_G(L) = \mu_G(R) = 1/2} \\
	& \geq \lambda \cdot \frac{1}{4} \cdot\frac{1}{2^2} \cdot \E_{i,j \sim L} \norm{u_i - u_j}^2 
		& \paren{ \paren{\max_{i \in L} \Delta(i)}/\paren{\min_{i \in L} \Delta(i)} \leq 2  } \mper 	
\end{align*}

\end{proof}

}{}

\newpage

\end{document}